\documentclass[twocolumn,floatfix,prb,aps,showpacs]{revtex4-1}
\usepackage{graphicx,amsmath,amssymb}

\newcommand{\bq}{\mathbf{q}}

\begin{document}

\title{Magnetoplasmons of the tilted-anisotropic Dirac cone material $\alpha-$(BEDT-TTF)$_2$I$_3$} 
\author{Judit S\'ari$^{1,2}$, Csaba T\H oke$^{2}$ and Mark O. Goerbig$^3$}
\affiliation{$^{1}$Institute of Physics, University of P\'ecs, H-7624 P\'ecs, Hungary}
\affiliation{$^{2}$BME-MTA Exotic Quantum Phases ``Lend\"ulet" Research Group, Budapest Univ. of Technology and Economics,
Institute of Physics, Budafoki \'ut 8., H-1111 Budapest, Hungary}
\affiliation{$^{3}$Laboratoire de Physique des Solides, CNRS UMR 8502, Univ. Paris-Sud, F-91405 Orsay Cedex, France}
\date{\today}

\begin{abstract}
We study the collective modes of a low-energy continuum model of the quasi-two-dimensional electron liquid
in a layer of the organic compound $\alpha-$(BEDT-TTF)$_2$I$_3$ in a perpendicular magnetic field.
As testified by zero magnetic field transport experiments and \textit{ab initio} theory,
this material hosts both massless and massive
low-energy carriers, the former being described by tilted and anisotropic Dirac cones.
The polarizability of these cones is anisotropic, and two sets of magnetoplasmon modes occur between any two
cyclotron resonances. We show that
the tilt of the cones causes a unique intervalley damping effect: the upper hybrid mode of one cone is
damped by the particle-hole continuum of the other cone in generic directions.
We analyse how the presence of massive carriers affects the response of the system,
and demonstrate how doping can tune $\alpha-$(BEDT-TTF)$_2$I$_3$ between regimes of isotropic and
anisotropic screening.
\end{abstract}


\maketitle

\section{Introduction}

The layered organic conductor $\alpha-$(BEDT-TTF)$_2$I$_3$,
an intensively investigated member of the (BEDT-TTF)$_2$I$_3$ family,\cite{Bender}
recently enjoys renewed interest due to the presence of Dirac cones in the low-energy band structure
under high hydrostatic pressure (above 15 kbar)\cite{pressure} or under uniaxial strain
(above 3 kbar along the $b$-axis).\cite{strain}
(For reviews, see Refs.~\onlinecite{oldreview}, \onlinecite{review},
\onlinecite{newreview} and \onlinecite{pressurereview}).
This material is a bulk crystal with a quasi-two-dimensional (2D) character,
as it consists of weakly coupled conductive BEDT-TTF [bis(ethylenedithio)-tetrathiafulvalene]
layers and insulating I$_3^-$ anion layers.

The existence of the gapless conical valleys at the Fermi energy under sufficiently high hydrostatic pressure or uniaxial
strain is testified by three experimental findings:
(i) The conductivity parallel to the layers is roughly constant from room temperature down to about 2 K,
while both the carrier density (measured via the Hall coefficient) and the mobility change by almost
six orders of magnitude in opposite
directions.\cite{pressure,strain,oldreview,review,Mori1984,Katayama2006,Kobayashi2007,Katayama2008,Katayama2009,Himura2011}
(ii) The conductivity perpendicular to the layers is greatly enhanced by a perpendicular magnetic field,
and shows a complex dependence on the parallel magnetic field component.\cite{Sugawara2010,Sato2011,Morinari2009,Osada2008}
(iii) The Shubnikov-de Haas oscillations and the integer quantum Hall effect have been observed recently.\cite{Tajima2013}
These observations have been interpreted in terms of spin- and valley-degenerate massless Dirac fermions,
assuming that only two exterior layers are hole-doped by the substrate.

Notice that none of the above observations are particularly sensitive to the geometry of the Dirac cones;
both the density of states and the Landau level spectrum are qualitatively identical for generic cones and upright
isotropic ones, which naturally arise in the low-energy spectrum of graphene\cite{graphene} and on the surface of
topological insulators.\cite{topins}
Our knowledge of the geometric properties of the conical valleys in $\alpha-$(BEDT-TTF)$_2$I$_3$ mainly
comes from theory, e.g., tight-binding modeling of the highest occupied molecular orbitals
at 3/4 filling,\cite{Mori1984,Katayama2006,Mori2010,Mori2013,footfilling}
extended Hubbard model calculations,\cite{Kobayashi2007,Katayama2008,Katayama2009,Himura2011}
and \textit{ab initio} band structure calculations.\cite{Kino,Ishibashi,Alemany}
On the other hand, theory also predicts\cite{Kino,Ishibashi,Alemany} a quadratic band maximum at
the $X$ point of the first Brillouin zone, and transport measurements by Monteverde \textit{et al.}\cite{Monteverde}
have identified carrier densities $\rho_L\approx2\times10^{8}$ cm$^{-2}$ (electrons) and
$\rho_Q\approx8\times10^{9}$ cm$^{-2}$ (holes) in the linear and the quadratic pockets, respectively,
which indicates an accidental overall hole-doping of their samples.
The presence of a hole-doped quadratic band maximum is, at first glance,
at odds with findings (iii) above,\cite{holedoping} and difficult to reconcile with (i),
although the smaller mobility of the massive carriers may be important in this connection.
Hence we believe a discussion of properties that can in principle probe the presence of massive carriers is timely.

Recent theoretical surveys by Mori\cite{Mori2010,Mori2013} have identified several other potential
massless Dirac materials among layered organic conductors.
In some of them, e.g., $\theta-$(BEDT-TTF)$_2$I$_3$, there might be a tiny gap.\cite{Miyagawa,Tajima2004}
In other cases, e.g., $\alpha$-(BEDT-TTF)$_2$KHg(SCN)$_4$, either the apex of the cones is far off the Fermi energy,
and the contact points give rise to electron-hole
pockets due to overtilt if the contact points approach the Fermi energy under pressure or strain,
or exhibit charge-density-wave structures that are incompatible with the zero-gap property in the
accessible range of the phase diagram.\cite{Mori2010}
One material, $\alpha$-(BEDT-TTF)$_2$NH$_4$Hg(SCN)$_4$ has no charge order and possesses Dirac points that
reach the Fermi energy under pressure,\cite{Choji2011}
but the linear bands become isotropic (like in graphene), making it less interesting from the point of
view of tilt-related electronic properties we are focusing on in this study.

Here, we study the particle-hole and collective excitations of a continuum model of the
quasi-2D electron gas in $\alpha-$(BEDT-TTF)$_2$I$_3$ in a magnetic field.
The electronic Coulomb interaction is taken into account within the random phase approximation (RPA).
Our continuum model includes both generic massless Dirac valleys and a massive one, such as those that are expected in the vicinity of
the Fermi energy from the topmost two bands in $\alpha-$(BEDT-TTF)$_2$I$_3$.
We restrict our attention to the low-energy and low-momentum features of the density-density response,
for which the structure of the highest two bands beyond these valleys is inessential.
To be specific, we will use the band parameters of $\alpha-$(BEDT-TTF)$_2$I$_3$ under high pressure from the literature,
and focus on the qualitative features that stem from the anisotropy and the tilt of the Dirac cones and the
presence of massive low-energy carriers.
Thus we extend the zero magnetic field analysis by Nishine \textit{et al.},\cite{Nishine}
who, neglecting the massive carriers, have found two plasmon modes,
and interpreted the appearance of the second mode as a consequence of plasmon filtering.\cite{filtering}

Our work generalizes analogous theoretical studies of plasmonic excitations in graphene.\cite{grafenBnull,grafenB,goerbigsummary}
In graphene in a perpendicular magnetic field, apart from the upper hybrid mode (UHM),
which is the zero-field  plasmon mode modified by the magnetic field,\cite{chiuplasmon} one can readily identify
linear magnetoplasmons, which run parallel to the boundary that separates the intra- and interband excitations
in the frequency-momentum plane.
Although the latter mode is essentially the coalescence of broadened interband excitations, its
huge spectral weight justifies its identification as a distinct collective mode of massless Dirac electrons. 
The UHM starts in the forbidden region at low momenta and merges into the linear magnetoplasmon mode in the interband
particle-hole excitation region.

We are particularly interested in the effects of the anisotropy and the tilt of the Dirac cones,
and include the massive holes to assess the relative importance of the two carrier types in the response.
Indeed, the valley degeneracy of the plasmonic modes and the UHM in the presence of a magnetic field
is lifted due to the tilt of the Dirac cones in opposite directions, in contrast to graphene, which has
isotropic and upright Dirac cones. Most saliently, the Coulomb coupling between the different
UHMs, which we investigate in the RPA, reveals a particular \textit{intervalley damping} effect that we 
discuss in contrast to a single-valley approximation, where Coulomb interaction
is taken into account only within a single cone and the plasmonic excitations can be associated to a particular valley.
The same mechanism must be present in other 2D systems with valleys of \textit{tilted} dispersion relations.

The article is structured as follows.
In Sec.~\ref{adatok} we review the low-energy band structure of $\alpha-$(BEDT-TTF)$_2$I$_3$,
introduce the three-valley model, and discuss the structure of the Landau states for all carriers.
In Sec.~\ref{interactions} we briefly comment on the interaction coupling in the distinct valleys.
In Sec.~\ref{methods} we describe the methods and approximations involved in the calculation of the
density-density response and the dielectric function.
Section \ref{results} discusses the results.
We provide a summary and discuss the possibly relevant experimental probes in Sec.~\ref{conclu}.
Details of the calculation are delegated to the Appendices.

\section{Continuum model for electrons in $\alpha-$(BEDT-TTF)$_2$I$_3$}

\label{adatok}

\begin{figure}[htbp]
\begin{center}
\includegraphics[width=\columnwidth, keepaspectratio]{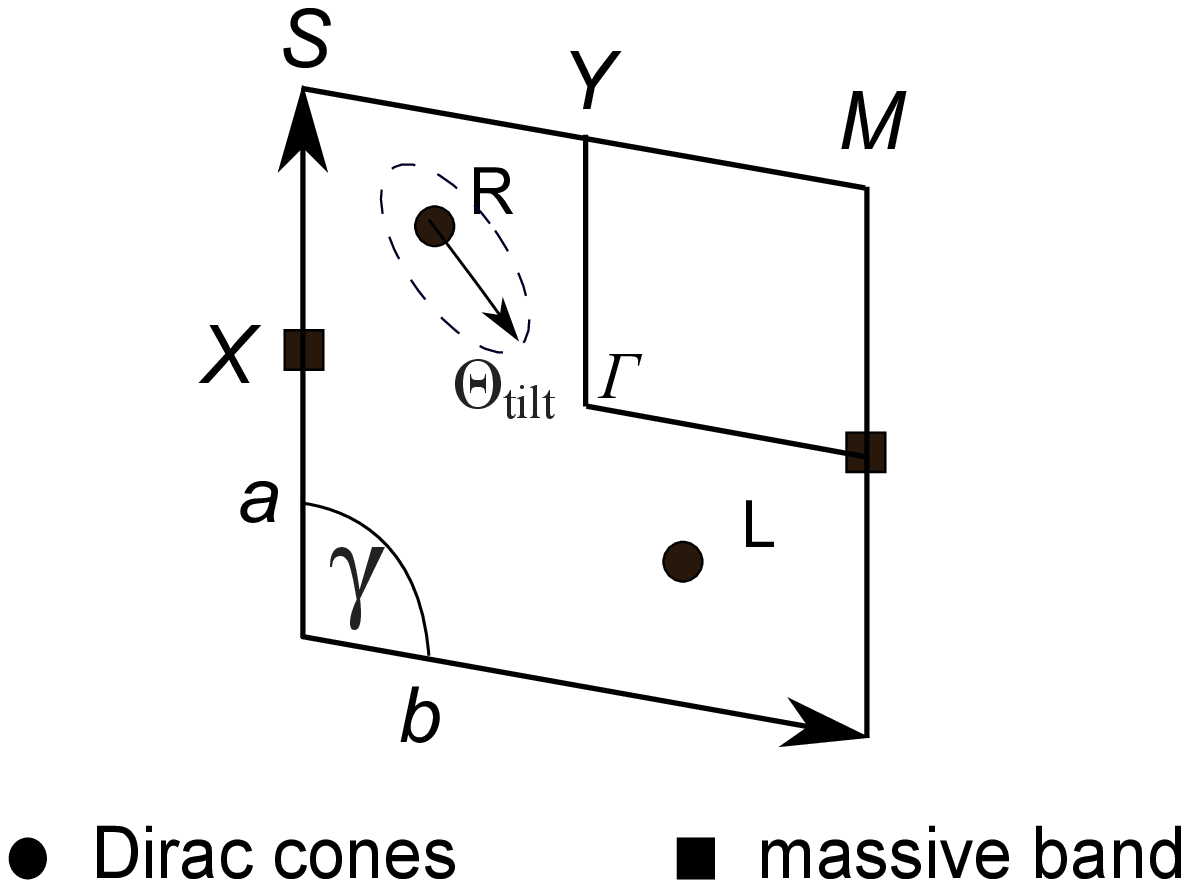}
\end{center}
\caption{\label{unitcell}
Schematic view of the first Brillouin zone of $\alpha-$(BEDT-TTF)$_2$I$_3$.
The high symmetry points, the location of the massive valley (squares),
and that of the Dirac cones (dots) are indicated.
The arrow from point $R$ indicates the smallest steepness of cone $R$.
Directions in the momentum plane will be related to this angle, $\theta_\text{tilt}$.
The dashed ellipse around $R$ represents an equipotential line. 
}
\end{figure}

$\alpha-$(BEDT-TTF)$_2$I$_3$ has a triclinic crystal structure
with strongly pressure-dependent lattice parameters.\cite{Kondo}
Regarding one layer as a 2D crystal, the unit cell is oblique,
with the primitive vectors enclosing an angle $\gamma\approx90.8^\circ$.
The length of the primitive vectors change monotonically from $a\approx9.1$ \AA\ and $b\approx10.8$ \AA\ to
$a\approx8.6$ \AA\ and $b\approx10.35$ \AA\ as the pressure is increased from ambient 
pressure to 17.6 kbar.\cite{Kondo}
See Fig.~\ref{unitcell} for a sketch of the first Brillouin zone with the distinguished points and
directions we refer to later.

The low-energy band structure can be derived from a tight-binding model that involves
the four relevant highest occupied molecular orbitals
of the four different BEDT-TTF molecules in the unit cell. Among the four bands,
only the upper two play a role as the filling is 3/4 at charge neutrality.\cite{footfilling}
These two bands may have contact points only in the absence of a stripe charge order.\cite{Mori2010}
At ambient pressure the stripe charge order occurs up to a phase transition around 135 K,\cite{transition}
but it is suppressed by pressure or strain\cite{pressure,strain,oldreview,review}.
In the non-stripe phase two tilted massless Dirac cones form in the vicinity of the contact points at
low-symmetry time-reversal related points $L$ and $R$ [see Fig.~\ref{unitcell}].
At even higher pressure or strain, the Dirac points coalesce at the $\Gamma$ point,\cite{merging} where a gap opens.
In addition to the linear valleys, there is a band maximum at the time-reversal symmetric point $X$
on the edge of the first Brillouin zone, which may host massive holes. 

In the present work we use a \textit{three-valley model}
that keeps a pair of generic massless Dirac cones of opposite tilt
and a massive valley with appropriate momentum/energy cutoffs.
Naturally, this approach is justified only as long as we study low-momentum and low-energy features.
We will show that this assumption is justified for our purposes because 
both the UHM and the linear magnetoplasmons occur within this range.

\subsection{The linear valleys}
\label{masslessmodel}

The massless Dirac carriers are suitably described by the minimal Weyl Hamiltonian\cite{Kobayashi2007,Goerbig} 
using four parameters:
\begin{equation}
\hat{H}_{\xi}(\mathbf{q}) = \xi \hbar \left(
\begin{array}{cc}
v_0^x q_x + v_0^y q_y  & v_x q_x - i \xi v_y q_y \\
v_x q_x + i \xi  v_y q_y  & v_0^x q_x + v_0^y q_y\\
\end{array} \right),
\label{Hamiltoni}
\end{equation}
where $\xi = +$ ($\xi = -$) represents the cone at $R$ ($L$).
This Hamiltonian obviously respects time-reversal 
symmetry, $\hat{H}_{+}(\mathbf{q})=\hat H^\ast_{-}( - \mathbf{q})$. 

Hamiltonian $\hat{H}_{\xi}$ has spin-degenerate bands, but the dispersions in valleys $\xi=\pm$ differ.
The inclination of the Dirac cone is determined by the combined effect of the tilt and the anisotropy. 
By the \textit{anisotropy} of the Dirac cone we mean the difference between $v_x$ and $v_y$, and we characterize it by
the parameter
\begin{equation}
\alpha = \sqrt{v_x/v_y}.
\end{equation}
By its \textit{tilt} we mean that the constant energy slices are not concentric because $(v_0^x,v_0^y)\neq(0,0)$.
For convenience, we will also use a rescaled and rotated coordinate system,\cite{Morinari2009}
defined by the transformation
\begin{equation}
\label{newcoord}
\left.
\begin{array}{l}
\tilde q_x = q_x\cos\theta + \frac{q_y}{\alpha^2}\sin\theta\\
\tilde q_y = -q_x\sin\theta  +  \frac{q_y}{\alpha^2}\cos\theta
\end{array}
\right\}.
\end{equation}
Rescaling the $q_y$ coordinate removes the anisotropy.
The rotation brings the $\tilde q_x$ coordinate in the tilt direction if we choose
\begin{equation}
\cos\theta=\cos\theta_\text{tilt}\equiv\frac{v_0^x v_y}{\sqrt{(v_0^y v_x)^2 + (v_0^x v_y)^2}}.
\end{equation}
After some straightforward algebra and a unitary transformation,\cite{Morinari2009} the Weyl Hamiltonian 
can be written as
\begin{equation}
\hat{H}_{\xi}(\tilde{q}_x,\tilde{q}_y) = \xi \hbar v_x \left(
\begin{array}{cc}
\eta \tilde{q}_x   &  \tilde{q}_x - i \xi \tilde{q}_y  \\
 \tilde{q}_x + i \xi \tilde{q}_y  & \eta \tilde{q}_x \\
\end{array} \right), \label{Hamiltoni_2}
\end{equation}
where we have introduced the dimensionless parameters
\begin{equation}
\eta = \sqrt{(v_0^x / v_x)^2 + (v_0^y /v_y)^2}\quad\text{and}\quad
\lambda = \sqrt{1-\eta^2}
\end{equation}
to quantify the tilt. 
Notice $0\le\eta\le1$, and that $\eta = 0$ corresponds to the case of graphene.

To be specific, we will use Kobayashi \textit{et al.}'s estimate of the velocity parameters\cite{Kobayashi2007}:
\begin{equation}
\left.\begin{array}{ll}
v_0^x = -9.4\times10^4\text{ m/s},&
v_0^y = -8.32\times10^4\text{ m/s},\\
v_x = 3.45\times10^5\text{ m/s},&
v_y = 2.45\times10^5\text{ m/s}
\end{array}\right\}
\label{velo}
\end{equation}
which yield $\eta = 0.437$, $\alpha = 1.18$, $\lambda = 0.89$, and $\theta_\text{tilt} = 51.14^{\circ}$. 
Various other velocity values are available in the literature,\cite{Himura2011,Katayama2008}
and they change considerably under pressure.\cite{Himura2011}
We emphasize that we will focus on qualitative features that hardly depend on this particular choice.

The Landau levels (LLs) of Weyl bands have been derived both semiclassically\cite{Goerbig}
and from a full quantum mechanical treatment.\cite{Morinari2009}
The spectrum is reminiscent of graphene, albeit with a reduced effective Fermi velocity:
\begin{equation}
\epsilon_{L,n} = \text{sgn}(n)\frac{\hbar}{\ell}\sqrt{2 v_x v_y \lambda^3 |n|}\label{landaulevel},
\end{equation}
where $\ell = \sqrt{\hbar/e B}$ is the magnetic length, $B$ is the applied magnetic field, and $n$ is an integer.
The corresponding orbirals in the Landau gauge are given in Appendix \ref{linorbit}.

We consider the linear approximation valid around each of the Dirac points separately in a circular region,
whose radius is chosen as about 1/8 of the side of the first Brillouin zone [$\approx 2\pi/(1\text{ nm})$],
consistently with the band structure obtained in Refs.~\onlinecite{Alemany},
\onlinecite{Kino} and \onlinecite{Katayama2008}.  
Using the velocities in Eq.~(\ref{velo}), the energy cutoff $E^c_L$
and the number of available LLs $n_{L}^{c}$ are
\begin{eqnarray}
E^c_L&\approx&0.16\text{ eV},\\
n_{L}^{c} &\approx&320/B\text{ [T]}.
\end{eqnarray}
Based on Monteverde \textit{et al.}'s electron density data,\cite{Monteverde}
the Fermi wave vector in the linear valleys is tiny,
$k^F_L =\sqrt{4\pi\rho_L/g_L}\approx2\times10^6$ m$^{-1}$, using $g_L=4$ for valley and spin degeneracy.
See Fig.~\ref{landaulevels} for a schematic view of the Landau level structure.

Notice that, while the massive valley breaks the particle-hole symmetry, the Landau levels of the linear 
valleys are particle-hole symmetric.
The restoration of particle-hole symmetry for the (tilted) massless carriers is due to the magnetic field, 
and it arises
because the cyclotron motion covers the complete isoenergy contours of the dispersion relation.
The effect of the tilt is therefore only to decrease the level
spacing by the factor $\lambda^{3/2}$ in Eq.~(\ref{landaulevel}).\cite{Morinari2009,Goerbig}

\subsection{The massive valley} 
\label{massivemodel}

The massive valley is centered around the $X$ point at the Brillouin zone edge (see Fig.~\ref{unitcell}).
It is taken as isotropic with an effective mass\cite{Monteverde}
\begin{equation}
\label{holemass}
m_Q\approx0.3 m_0
\end{equation}
in terms of the free electron mass $m_0$.
The valley is hole-like, a paraboloid of revolution open from below.
The Hamiltonian of the massive band is
\begin{equation}
\hat{H}_Q  = E_\text{offset}-\frac{(-i\hbar\nabla + e \mathbf{A})^2}{2 m_Q},
\end{equation}
and the Landau level spectrum is given by
\begin{equation}
\label{quadenergy}
\epsilon_{Q,n}=E_\text{offset}-\hbar\omega_c\left(n + \frac{1}{2}\right),
\end{equation}
where $\omega_c=eB/m_Q$ is the appropriate cyclotron frequency and $n\ge0$ is an integer.
(Notice that these nonnegative integers actually number hole LLs.)
Combining the carrier densities measured by Monteverde \textit{et al.}\cite{Monteverde} with
Eqs.~(\ref{velo}) and (\ref{holemass}), the top of the massive band is about
\begin{equation}
E_\text{offset}\approx0.46\text{ meV}
\end{equation}
above the Dirac point of the massless valleys.
The corresponding eigenstates in the Landau gauge $\mathbf{A} = (-y B ,0,0)$ are
\begin{equation}
\zeta_{n,q}(\mathbf{r})=
\frac{e^{iqx}}{\sqrt{\sqrt{\pi}2^n n!\ell2\pi}}
e^{-\frac{1}{2}\left(\frac{y}{\ell}-q\ell\right)^2}
H_n\left(\frac{y}{\ell}-q\ell\right),
\label{masseigenstates}
\end{equation}
where $H_n(x)$ is a Hermite polynomial.

The quadratic approximation is valid in a circle around point $X$ in momentum space,
with a radius that is estimated as 17.5 \% of the side of the first Brillouin zone.
This yields the cutoff energy $E^c_Q$ and the number of LLs is $n_Q^{c}$ as
\begin{eqnarray}
E^c_Q&\approx&0.15\text{ eV},\\
n_Q^{c}&\approx&390/B\text{ [T]}.
\end{eqnarray}
The Fermi mometum $k_Q^F$ of the massive band\cite{Monteverde}
is $k_Q^{F} = \sqrt{4\pi\rho_Q/g_Q}\approx2\times10^7$ m$^{-1}$, using $g_Q = 2$ for spin degeneracy.

\section{The interaction strength}

\label{interactions}

The relative strength of the interaction for each carrier type is characterized by
the ratio between the interaction energy scale $E_\text{int} = e^2/(4\pi\epsilon_0\epsilon_r l)$
and the kinetic energy scale $E_\text{kin}$ at the characteristic length scale $l \approx 1/ k^F$.
Here, $\epsilon_r$ is the relative dielectric constant of the material.
The kinetic energy scale depends on the carrier type.

For the massive carriers, $E_\text{kin}\propto l^{-2}$,  and the ratio $r_s$ depends on the Fermi wave vector,
\begin{equation}
r_s=\frac{m_Q}{a_0\epsilon_r m_0k_Q^{F}}\approx\frac{300}{\epsilon_r},
\end{equation}
where $a_0$ is the Bohr radius.
For the last number we have used the Fermi momentum as estimated for a specific sample in Subsec.~\ref{massivemodel}. This ratio is traditionally called the Wigner-Seitz radius.

In contrast to the parabolic bands, the kinetic energy of massless Dirac carriers
scales in the same manner as the interaction energy, $E_\text{int}\propto l^{-1}$, hence there is no
characteristic length such as the Bohr radius. Indeed, the ratio between interaction and kinetic energy
is independent of the electron density, and it may be characterized by a ``fine structure constant
of $\alpha-$(BEDT-TTF)$_2$I$_3$''
\begin{equation}
\alpha_{\alpha-\text{(BEDT-TTF)}_2\text{I}_3}=\frac{\alpha c}{\sqrt{v_xv_y}\epsilon_r} \approx \frac{20}{\epsilon_r}, 
\end{equation}
where $\alpha\approx1/137$ is the fine structure constant of quantum electrodynamics,
and $c$ is the speed of light.
Notice that the average Fermi velocity $\sqrt{v_xv_y}\approx 10^5$ m/s is an order of magnitude smaller than the
corresponding velocity in graphene.\cite{graphene}
This is the origin of the rather large value of the ratio between the interaction and the kinetic energy.

In view of this high value of the energy ratio for both the massless and massive carriers
in $\alpha-$(BEDT-TTF)$_2$I$_3$, one may expect the formation of correlated phases, such as the Wigner crystal.
For the conventional 2D electron gas (2DEG), $r_s \gtrsim 37$ is required
to reach the Wigner crystal phase of the massive carriers.\cite{Wignercrystal}
This would require in turn a dielectric constant $\epsilon_r<10$. 
To the best of our knowledge there is no available experimental value for $\epsilon_r$ in $\alpha-$(BEDT-TTF)$_2$I$_3$,
but we expect it to be such as to rule out the Wigner crystal.
In the linear valleys, the high value of the dielectric constant compensates
for the small Fermi velocity, in which case, just like in graphene,
one would not expect an instability of the semimetalic phase.\cite{KotovDrut}
Throughout the article we assume that the Wigner crystal can be discarded, and use $\epsilon_r = 10$.

\section{Methods}
\label{methods}

We identify the collective modes in the random phase approximation.
Assuming that all but the topmost two bands are inert, the density-density response
$\chi^\text{RPA}(\mathbf{q}, \omega)$ is determined by the bare polarizability $\chi^{(0)}(\mathbf{q},\omega)$ as
\begin{gather}
\label{chiRPA}
\chi^\text{RPA}(\mathbf{q}, \omega) =  \frac{\chi^{(0)} (\mathbf{q}, \omega)}{\epsilon^\text{RPA} (\mathbf{q}, \omega)},\\
\label{dielectricfunc}
\epsilon^\text{RPA}(\mathbf{q},\omega) = 1 - v(\mathbf{q}) \ \chi^{(0)} (\mathbf{q}, \omega),\\
\chi^{(0)}(\mathbf{q}, \omega)=
-i\text{Tr}\int\frac{\mathrm{d}E}{2\pi}\int\frac{\mathrm{d}^2 \mathbf{p}}{(2\pi)^2}
\mathcal{G}^{(0)}\left( \mathbf{p}, E\right)\times\nonumber\\ 
\times\mathcal{G}^{(0)}\left( \mathbf{p}+\mathbf{q}, E+\omega\right).
\label{chidef}
\end{gather}
Here $v(\mathbf{q}) = e^2/2\epsilon_r\epsilon_0q$ is the bare Coulomb interaction,
and the bare propagator $\mathcal{G}^{(0)}\left(\mathbf{p},E\right)$ is a $4\times4$ matrix in the original lattice model
related to the amplitudes on the four BEDT-TTF molecules in the primitive cell.
Reliable band structure information, however, is only available near the contact points of the topmost two bands.
Hence we use the three-valley model introduced in Sec.~\ref{adatok}, and approximate the polarizability accordingly.

In Eq.~(\ref{chidef}), $\chi^{(0)}(\mathbf{q},\omega)$ picks up a contribution only if $(\mathbf p,E)$
and $(\mathbf q+\mathbf p,\omega+E)$ specify one filled and one empty state.
As we will restrict our attention to $|\omega|<\min(E_Q^c,E_L^c)\approx0.15$ eV, we can ignore the
cases when both of these points are outside the vicinities of the $L$, $R$ and $X$ points, respectively,
assuming that the Fermi energy is near the contact points.
We can also ignore the cases where the states $(\mathbf p,E)$ and $(\mathbf q+\mathbf p,\omega+E)$ are
in distinct valleys, as long as we focus on small momenta $|\mathbf q|<K\equiv\min(k_Q^c,k_L^c)$.
This approximation is justified because the Coulomb interaction intervenes in 
the dielectric function $\epsilon^\text{RPA}(\mathbf{q},\omega)$, and
suppresses intervalley contributions in comparison to intravalley contributions at fixed $\omega$
by a factor $v(K)/v(k_F)=k_F/K\ll 1$ near the characteristic Fermi momentum $k_F$.
Furthermore, for small doping $|\mu|\ll\min(E_Q^c,E_L^c)$, we can also ignore the transitions that involve a state
in the range of validity of the linear/quadratic approximations 
(Subsecs.~\ref{masslessmodel} and \ref{massivemodel}) and a state outside of this domain.

Thus we can safely approximate the bare polarizability for our limited purposes as the sum of the contributions
of intravalley particle-hole pairs:
\begin{equation}
\label{chi0teljes}
\chi^{(0)}(\mathbf{q},\omega)\approx
\chi^{(0)}_{L}(\mathbf{q},\omega)+\chi^{(0)}_{R}(\mathbf{q},\omega)+\chi^{(0)}_{Q}(\mathbf{q},\omega),
\end{equation}
where $\chi^{(0)}_\text{V}(\mathbf{q},\omega)$ is the polarizability contribution from intravalley transitions
in valley $V$, with $V=L,R,Q$.
Further, when studying features of a wave length much larger than the unit cell, the distribution of
amplitudes in the unit cell can be ignored.
Using the wave functions in Eq.~(\ref{masseigenstates})
for massive carriers, their contribution is\cite{2DEGchi}
\begin{multline}
\chi^{(0)}_{Q}(\mathbf{q}, \omega)=
\frac{1}{2\pi \ell^2} \sum_{n'\le n^F_Q} \sum_{n>n^F_Q}\\
\left( \frac{ | F_{n,n'}( \mathbf{q})| ^2  }{\omega - \epsilon_n +\epsilon_{n'} +i \delta}
+\frac{  |F_{n',n}( \mathbf{q})| ^2  }{-\omega - \epsilon_n +\epsilon_{n'} - i \delta }\right),
\label{chi0mass}
\end{multline}
where we have used the form factors of the 2DEG,
\begin{equation}
F_{n',n}(\mathbf{q})= \sqrt{\frac{n!}{n'!}}\left(\frac{q_x-iq_y}{\sqrt{2}}\ell\right)^{n'-n}
L_{n}^{n'-n}\left(\frac{|q|^{2}\ell^2}{2}\right)e^{-\frac{|q|^2 \ell^2}{4}}
\end{equation}
for $n'\geq n$, whereas for $n'<n$,
\begin{equation}
F_{n,n'}(\mathbf{q})= F^\ast_{n',n}(-\mathbf{q}).
\end{equation}
Here $L_n^m(z)$ denotes an associated Laguerre polynomial.
Similarly, using the wave functions in Eqs.~(\ref{nagypszi}-\ref{nagypszivege}) for massless carriers,
the contribution from valley $\xi=\pm$ is
\begin{multline}
\chi^{(0)}_{\xi} (\tilde{q}_x, \tilde{q}_y, \omega)=
\frac{1}{2 \pi \alpha^2  \ell^2} \sum_{n'\le n^F_L} \sum_{n>n^F_L}\\
\left( \frac{|\mathcal{F}^{\xi}_{n,n'}(\tilde{q}_x, \tilde{q}_y )| ^2  }{\omega - \epsilon_n +\epsilon_{n'} +i \delta}
+\frac{|\mathcal{F}^{\xi}_{n',n}( \tilde{q}_x, \tilde{q}_y )| ^2  }{-\omega - \epsilon_n +\epsilon_{n'} - i \delta }\right).
\label{chi0lin}
\end{multline}
The corresponding form factors $\mathcal{F}^{\xi}_{n,n'}(\tilde{q}_x, \tilde{q}_y)$ are
defined in Appendix \ref{applinpol}.
Using the expressions of the form factors, one can check
that $\chi^{(0)}_{L}(\mathbf{q}, \omega)$ and $\chi^{(0)}_{R}(\mathbf{q},\omega)$ are related
by the change of the sign of the tilt $\eta$.

For illustration purposes, we consider occasionally 
the polarizability, the dielectric function and the density-density response functions
stemming from only one or two valleys, even if such model systems are unphysical.
Thus $\epsilon^\text{RPA}_V(\mathbf{q},\omega)$ and $\chi^\text{RPA}_V(\mathbf{q},\omega)$ are defined
in terms of $\chi^{(0)}_V(\mathbf{q},\omega)$ in an obvious manner, where $V$ is either $L$, $R$,
or $Q$ for the respective valley.
Moreover, we also consider cases with only two valleys taken into account.
Sometimes this is a physically relevant situation, e.g., when the system is electron-doped or charge-neutral
and the magnetic field is large, sometimes a theoretical contrast.
Then we define
$\chi^{(0)}_{V_1+V_2}(\mathbf{q},\omega)=\chi^{(0)}_{V_1}(\mathbf{q},\omega)+\chi^{(0)}_{V_2}(\mathbf{q},\omega)$
and again $\epsilon^\text{RPA}_{V_1+V_2}(\mathbf{q},\omega)$ and $\chi^\text{RPA}_{V_1+V_2}(\mathbf{q},\omega)$ follow
in analogy to Eqs.~(\ref{chiRPA}) and (\ref{dielectricfunc}).

\section{Results and Discussion}
\label{results}

The particle-hole excitation spectrum (PHES) and the collective modes of $\alpha-$(BEDT-TTF)$_2$I$_3$
at low energies are determined by the two massless Dirac cones and the massive hole pocket.
The way the three valleys contribute to the density-density response of a layer
depends on the doping and the perpendicular magnetic field.
For significant electron doping the massive band is full at zero temperature and only the massless carriers
contribute to the transport (c.f.\ Subsec.~\ref{edoping}).

\begin{figure}[htbp]
\begin{center}
\includegraphics[width=\columnwidth, keepaspectratio]{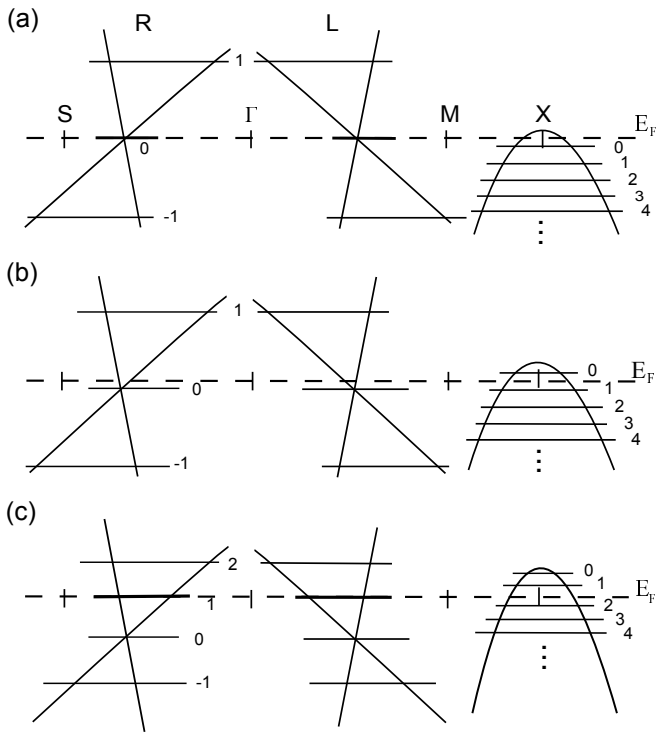}
\end{center}
\caption{\label{landaulevels}
Schematic view of the Landau level structure of the two Dirac cones and the massive valley,
with the chemical potential at charge neutrality indicated.
(a) $B>B_{00}\approx 2.5$ T, (b) $B_{11}<B<B_{00}$, and (c) $B\lesssim B_{11}\approx 0.06$ T.
}
\end{figure}

At charge neutrality, the chemical potential is between the Dirac point and the top of the massive band
in zero magnetic field.
By turning on the magnetic field, the central four-fold degenerate $n=0$ LL of the massless valleys is
fixed at the Dirac points, but the energy $E_\text{offset}-\hbar\omega_c/2$ of the topmost $n=0$ LL
of the massive valley decreases.
Let $B_{nm}$ be the field when the $m$-th LL of the massive valley coincides with the $n$-th LL of the conical valleys.
For $B>B_{00}$, the $n=0$ LL of the Dirac valleys is half-filled and the
completely electron-filled massive valley is inert, c.f.\ Fig.~\ref{landaulevels}(a).
The response is entirely due to the massless valleys, similarly to the electron-doped case (Subsec.~\ref{edoping}).
In the interval $B_{11}<B<B_{00}$ the chemical potential lies between the $n=0$ LL of the linear band and
the $n=0$ LL of the massive band ($n_L^F=0$, $n_Q^F=1$), c.f.\ Fig.~\ref{landaulevels}(b).
The excess electrons in the massless valleys (two per flux quanta)
exactly compensate for the excess holes in the massive pocket.
Now all valleys contribute to the density-density response (Subsec.~\ref{threevalleys}).
For $B\lesssim B_{11}$, the $n=1$ LL of the massive valley becomes empty and the four-fold degenerate
$n=1$ LL of the massless Dirac valleys is half-filled; the chemical potential is set somewhere in the
(naturally broadened) $n=1$ LL of the massless Dirac fermions, c.f.\ Fig.~\ref{landaulevels}(c).
Combining Refs.~\onlinecite{Monteverde} and \onlinecite{Kobayashi2007},
we estimate $B_{11}\approx0.06$ T and $B_{00}\approx2.5$ T.

In significantly hole-doped samples all valleys contribute.

\subsection{The response of massless carriers}
\label{edoping}

With a significant electron-doping in mind, we set $n_L^F = 2$, for which $n^F_Q=0$, i.e.,
the massive band is full and inert for all realistic values of the magnetic field $B$.

\subsubsection{The response of a single cone}

\label{sec:SCA}

To highlight the direction-dependent effects, we first consider the response of a single tilted massless Dirac cone,
as described by the Weyl Hamiltonian in Eqs.~(\ref{Hamiltoni}) and (\ref{Hamiltoni_2}).
Although this model is not directly related to a concrete physical situation, it reveals
some basic phenomena associated with the tilt of the Dirac cones, 
such as the direction-dependent plasmonic dispersions and the characterization
of the different regions of allowed and forbidden particle-hole excitations.
This preliminary analysis within the single-cone approximation thus helps us understand
the effect of the Coulomb coupling between the two different valleys, discussed in Sec.~\ref{bothcones}.

Regarding the possibility of particle-hole excitations, the $(\omega,q)$ plane can be divided into several regions.
Here we follow the notation Nishine \textit{et al.}\cite{Nishine} have introduced for the PHES
in zero magnetic field.
The regions of possible intraband (A-region) and interband (B-region) particle-hole excitations
are separated by the boundary line $\omega_{res}$.
Due to the opposite tilt, the various regions and their boundaries differ for the two cones in any general direction.
Both regions are divided into subregions, as explained below.

\begin{figure}[htbp]
\begin{center}
\includegraphics[width=\columnwidth, keepaspectratio]{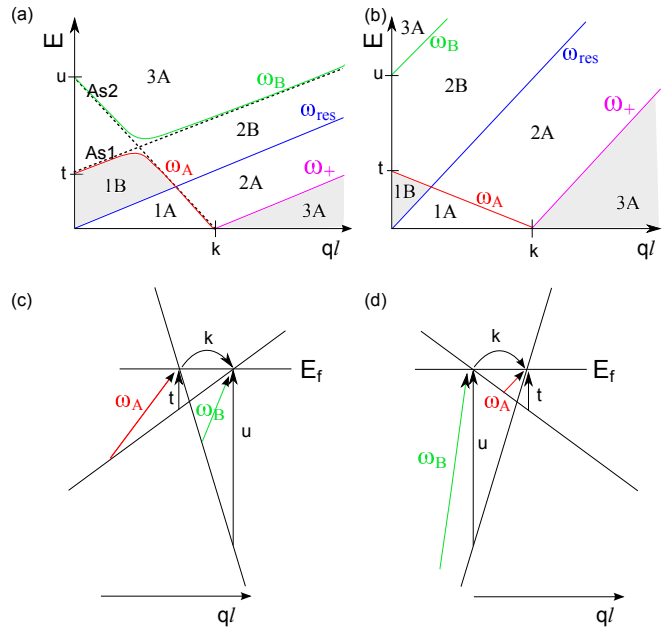}
\end{center}
\caption{\label{nishinetiltdirection} 
(Color online)
The regions and subregions of the $(\omega,q)$ plane from the point of view of (a) cone $R$, and (b) cone $L$
in a particular direction $\theta=\theta_\text{tilt}$.
(c,d) Cuts of cones $R$ and $L$ in the same direction.
In the direction $\theta=\theta_\text{tilt}+\pi$, the cones $L$ and $R$ are interchanged, such that the
panels (a,c) would correspond to cone $L$ and (b,d) to cone $R$.
The distinguished energies $t,u$, momentum $k$, and asymptotes $As1,As2$ are indicated. The gray shading indicates the forbidden
regions 1B and 3A, where there are no particle-hole excitations in the absence of electron-electron interactions.
}
\end{figure}

Figures \ref{nishinetiltdirection}(a) and (b) 
depict the $(\omega,q)$ plane in the direction $\theta=\theta_\text{tilt}$,
where $\theta=\arctan(q_y/q_x)$ is the angle of the momentum.
In this direction the steepness of cone $R$ is minimal, while that of cone $L$ is maximal.

First consider cone $R$ in this direction [Figs \ref{nishinetiltdirection}(a) and (c)].
Zero-energy excitations must have a momentum transfer less than the major axis $k$ of the ellipsoidal Fermi surface.
Excitations with higher momenta require a minimal energy, which defines the upper boundary $\omega_+$ of the forbidden
subregion 3A [c.f.\ Fig \ref{nishinetiltdirection}(a,b)], where $\text{Im}\chi_R^{(0)} = 0$.

Excitations with zero momentum transfer must be interband.
For $q=0$, there are two special energies.
The smallest one is denoted $t$ in Fig.~\ref{nishinetiltdirection}.
Below this energy no $q=0$ excitation is possible.
If we increase the total momentum of the particle-hole pair in the positive manner,
we must move the hole in the $-q$ direction [see arrow at $\omega_A$ in Fig \ref{nishinetiltdirection}(c)].
Then the excitation energy increases, which explains the rise of the boundary line $\omega_A$ for small momenta.
By thus decreasing the hole's wave vector, one follows the asymptote $As1$ in Fig \ref{nishinetiltdirection}(a),
which ultimately merges into the boundary line $\omega_B$.

Starting from $u$, one can decrease the excitation energy with positive total momentum;
this branch corresponds to $\omega_B$ and follows asymptote $As2$ at small momenta, which merges into $\omega_A$ in the intraband region. 
At larger values of the momentum, $\omega_B$ asymptotically follows $As1$.
The reason why the boundaries $\omega_A$ and $\omega_B$ deviate from the asymptotes $As1,As2$ and
do not intersect each other is due to the 2D nature of the excitations, i.e.,
one needs to consider transitions outside of the one-dimensional cut examined so far.  
Between $\omega_A$ and $\omega_{res}$ no particle-hole excitations exist,
thus region 1B is forbidden, just like 3A for intraband excitations.

If one considers excitations of cone $L$ in the same momentum direction,
the shapes of boundaries $\omega_B$ and $\omega_A$ differ considerably [Fig \ref{nishinetiltdirection}(b)].
The smallest energy of a $q=0$ excitation is still $t$. However, in contrast to cone $R$, 
it is possible to decrease the energy from $t$ by moving the hole in the $-q$ direction [Fig \ref{nishinetiltdirection}(d)], and
the corresponding branch defines boundary $\omega_A$.
Starting from $u$ one can only increase the excitation energy with a negative total momentum, and one thus obtains
the branch corresponding to $\omega_B$.
Naturally, the two cones are related by time-reversal symmetry, which also changes the sign of $\mathbf q$.
The role of the $R$ and $L$ cones are therefore interchanged in the $\theta_\text{tilt}+\pi$ direction, and 
the regions of cone $R$ ($L$) are then those in
Fig.~\ref{nishinetiltdirection}(b,d) [Fig.~\ref{nishinetiltdirection}(a,c)].

\begin{figure}[htbp]
\begin{center}
\includegraphics[width=\columnwidth, keepaspectratio]{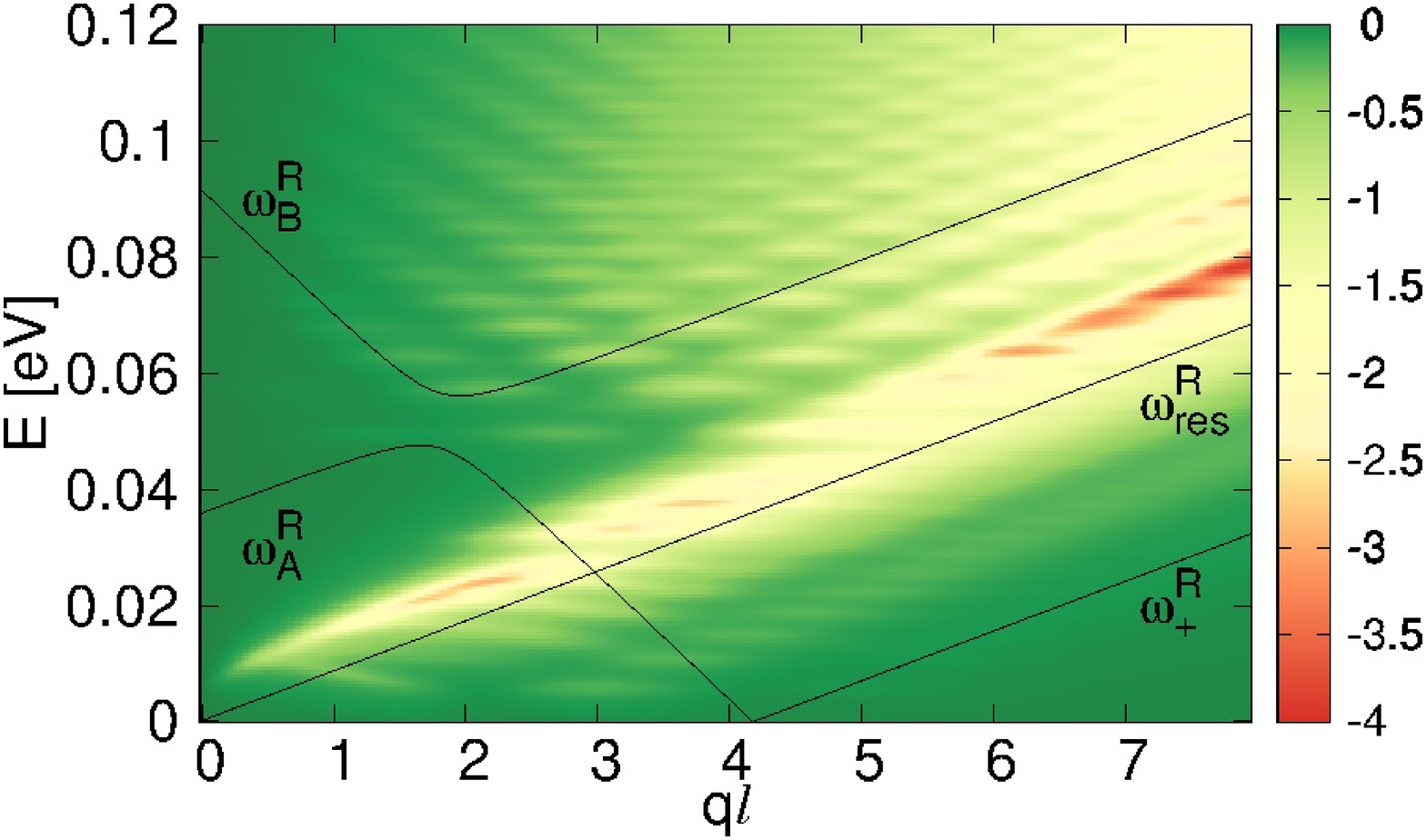}

\includegraphics[width=\columnwidth, keepaspectratio]{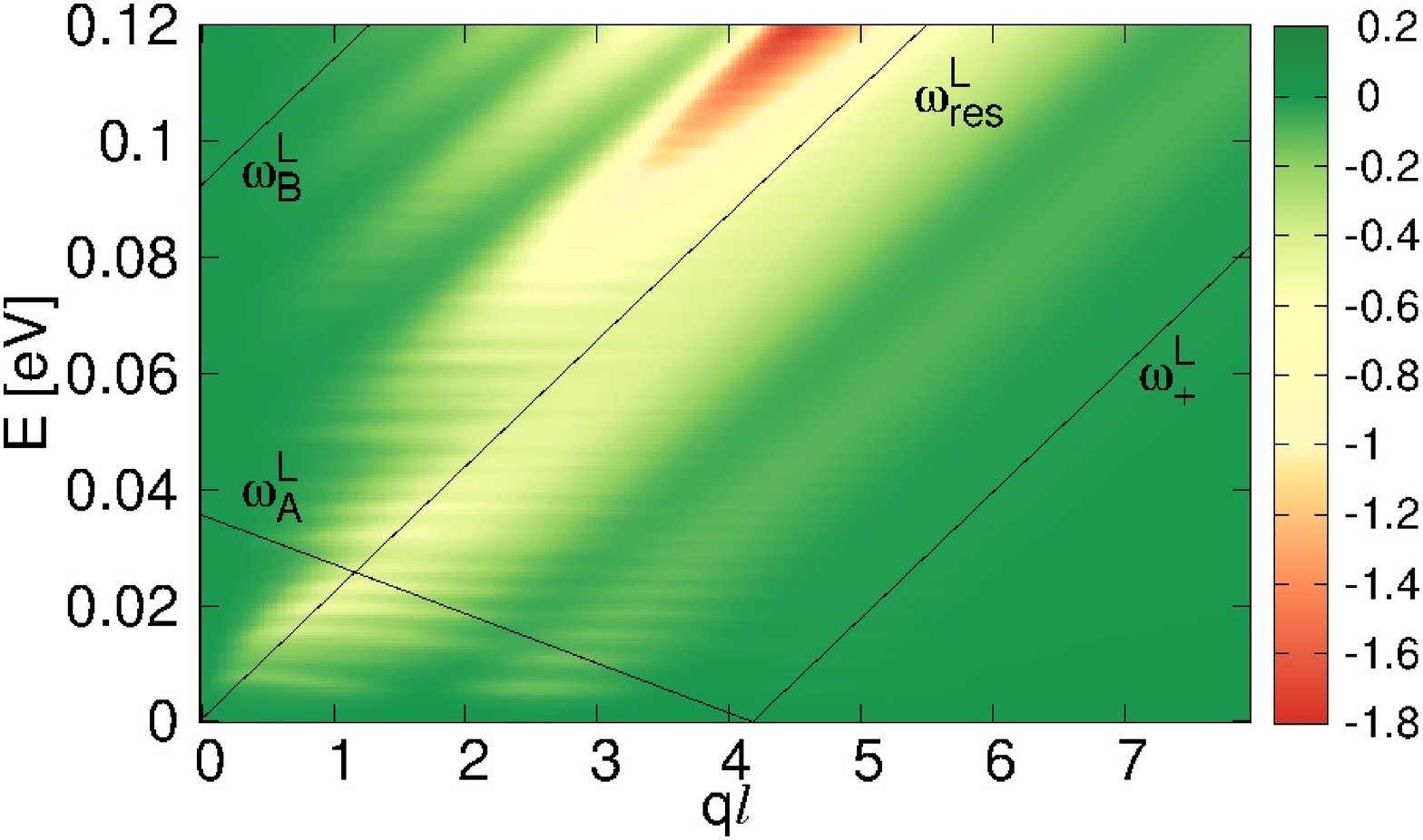}

\includegraphics[width=\columnwidth, keepaspectratio]{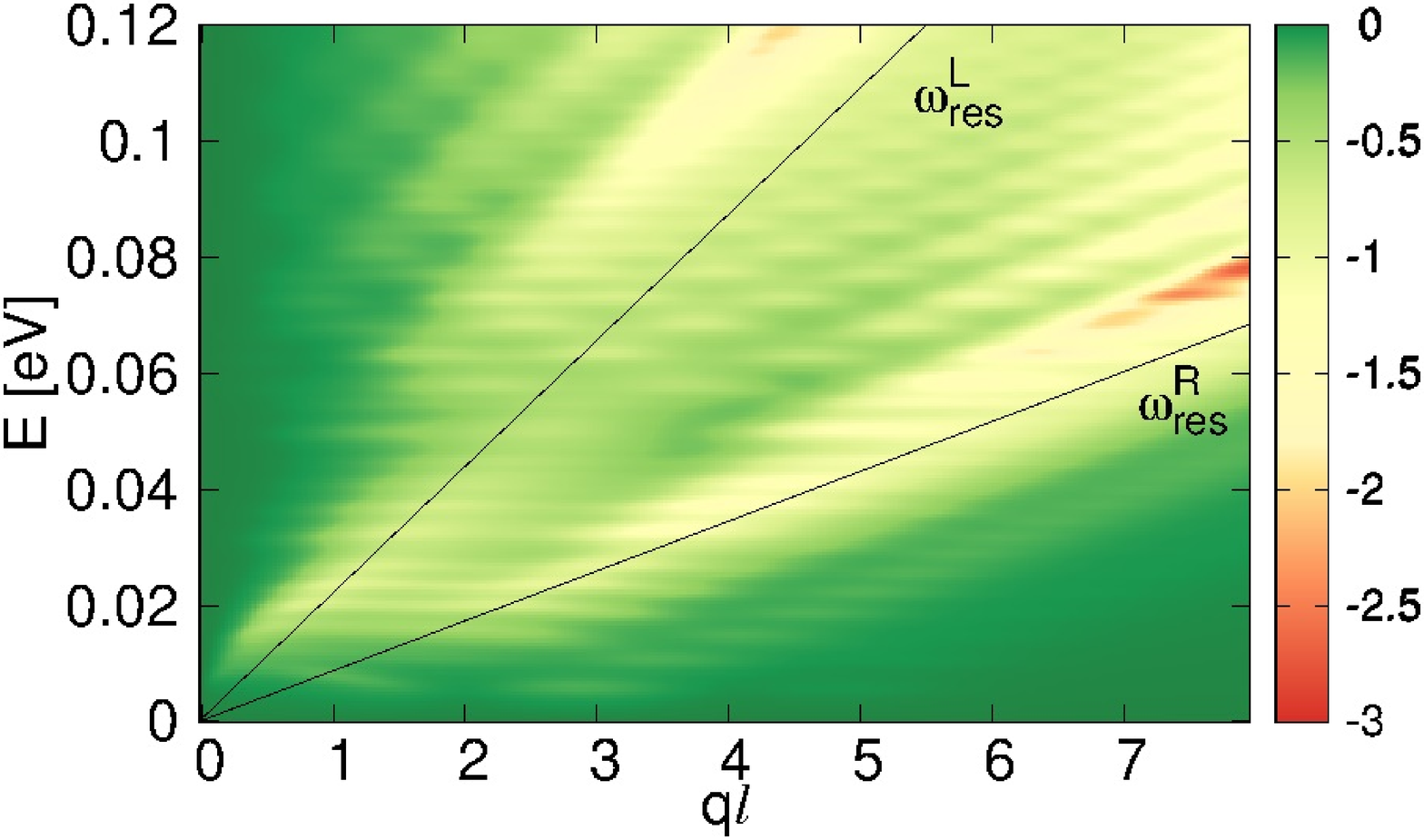}
\end{center}
\caption{
(Color online)
The imaginary part of the density-density response of massless carriers,
divided by the density of states at the Fermi energy.
The topmost filled Dirac Landau level is $n_L^F = 2$, the other parameters are $B = 4$ T and $\epsilon_r=10$.
The first two panels consider the cones individually;
(a) shows $\text{Im}\chi^\text{RPA}_{R}$ in the direction of its maximal tilt $\theta_\text{tilt}$
(or $\text{Im}\chi^\text{RPA}_{L}$ in the direction $\theta_\text{tilt}+\pi$), and
(b) $\text{Im}\chi^\text{RPA}_{L}$ in the direction $\theta_\text{tilt}$
(or $\text{Im}\chi^\text{RPA}_{R}$ in the direction $\theta_\text{tilt}+\pi$).
Panel (c) shows $\text{Im}\chi^\text{RPA}_{L+R}$ in direction $\theta_\text{tilt}$,
which is the response of the total system for electron doping, or for $B>B_{00}\approx 2.5$ T at charge neutrality.
The straight lines are the boundaries of regions relevant at $B=0$; c.f.\ Fig.~\ref{nishinetiltdirection}(a,b).
Notice that in panel (c), we have only depicted the lines $\omega_{res}^{R/L}$ for the two cones, as a guide to the eye.
}
\label{egydirection1}  
\end{figure}

These zero-field considerations correctly describe the regions of allowed transitions also for non-zero
magnetic fields since the wave vector in the PHES is that of (neutral) electron-hole pairs. It therefore
remains a good quantum number also in the presence of a magnetic field.
Figures \ref{egydirection1}(a) and (b) show $\text{Im}\chi^\text{RPA}_{R}$ and $\text{Im}\chi^\text{RPA}_{L}$
in the direction of $R$'s smallest velocity $\theta=\theta_\text{tilt}$, where we have also plotted the boundaries 
$\omega_+^{R/L}$, $\omega_A^{R/L}$, $\omega_B^{R/L}$, and $\omega_{res}^{R/L}$, the latter corresponding to the frequency $\omega_{res}$
for the right and the left cone, respectively. 
Notice that we consider a finite broadening  (inverse quasiparticle lifetime)
$\delta=0.1\hbar\sqrt{v_xv_y}/\ell$. Whereas
treating $\delta$ as an energy-independent constant is a crude approximation,\cite{footpos}
we use it here only as a phenomenological parameter that renders the structure of the PHES more
visible.

\begin{figure}[htbp]
\begin{center}
\includegraphics[width=\columnwidth, keepaspectratio]{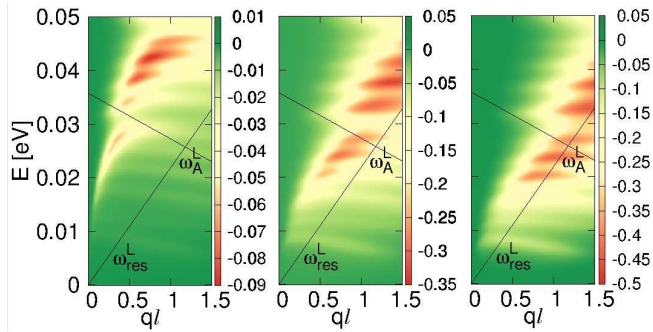}
\end{center}
\caption{
(Color online)
The dependence of the upper hybrid mode of massless Dirac fermions on the background dielectric constant $\epsilon_r$.
We show $\text{Im}\chi^\text{RPA}_{L}$ in the direction opposite to cone $L$'s maximal tilt,
$\theta=\theta_\text{tilt}$, for (a) $\epsilon_r=1$, (b) $\epsilon_r=4$ and (c) $\epsilon_r=7$.
The physical parameters are the same as in Fig.~\ref{egydirection1}, whose panel (b) depicts $\epsilon_r=10$.
}
\label{epsilonfig}
\end{figure}

As shown in Fig. \ref{egydirection1}(a),
the UHM is present with a considerable spectral weight in the originally forbidden region 1B
of cone $R$ in its tilting direction $\theta=\theta_\text{tilt}$.
For cone $L$, on the other hand, the spectral weight in region 1B in the same direction is definitely smaller
[Fig. \ref{egydirection1}(b)]. As shown in Fig.~\ref{epsilonfig}, where we have plotted a zoom of 
$\text{Im}\chi^\text{RPA}_{L}$ in region 1B, the spectral weight could nevertheless be increased considerably
by lowering the dielectric constant $\epsilon_r$. However, a low spectral weight is expected in the physically
relevant situation, for an estimate $\epsilon_r\approx10$ (c.f.\ Sec.~\ref{interactions} above).
The reason is that the weakening of the bare interaction pushes the zeros of the RPA dielectric function
to higher momenta at fixed energy; eventually the UHM is forced to the borderline of region 1B for
cone $L$ in the given direction.
For cone $R$ the tendency is similar but less significant, c.f.\ Fig.~\ref{egydirection1}(a).

After leaving region 1B, the UHM merges into the linear magnetoplasmon mode in region 2B.  
The concentration of the spectral weight near $\omega_{res}$ is in accordance with the $B=0$
limit\cite{Nishine} and the $B\neq0$ behavior of graphene.\cite{goerbigsummary}
Recall that the UHM arises due to the modification of the classical plasmons by
cyclotron motion.\cite{chiuplasmon} It does not require interband excitations,
thus it forms by a transfer of spectral weight from the intraband PHES to the originally forbidden region 1B.

The regular ``island'' structures, i.e., the alternation of parts with high and low spectral weight in the 
$(\omega,q)$-plane, which is observable at high energies in Fig.~\ref{egydirection1} and
subsequent figures, evolve from interband particle-hole excitations as the magnetic field is turned on.
Their structure in momentum space is related to the structure of high-index Landau orbitals involved and
it is exactly of the same origin as in the case of nontilted cones: 
the form factors in Eqs.~(\ref{form1}) to (\ref{form2}) contain Laguerre polynomials with many zeros for
transitions between deep-lying and high-energy Landau levels.\cite{grafenB}

To summarize, the density-density response of a generic massless Dirac cone in a perpedicular magnetic
field exhibits an upper hybrid mode and linear magnetoplasmons with anisotropic velocities,
as a plausible generalization of the graphene case.\cite{goerbigsummary}

\subsubsection{Both cones considered}
\label{bothcones}

Fig.~\ref{egydirection1}(c) shows $\text{Im}\chi^\text{RPA}_{L+R}$ for both cones
in a fixed direction $\theta=\theta_\text{tilt}$ of the momentum plane.
Notice that for the electron-doped case this is actually the total density-density response.

The UHM of cone $R$ disappears, though it was the most dominant part of the response in the single-cone approximation. 
The UHM of cone $L$ is still present, though with a reduced spectral weight in its own forbidden region.
Its linear magnetoplasmon mode manifests itself, although at high energies ($R$'s region 2B) it is surrounded by
the interband particle-hole excitations of cone $R$, which have a modest spectral weight in the single cone approximation.
Both modes are approximately in the same place as they were when only one cone was considered.
For interpretation, compare Fig.~\ref{nishinetiltdirection}(c),
where we sketch the $(\omega,q)$ plane in a fixed direction $\theta=\theta_\text{tilt}$.
The forbidden region of cone $L$ lies entirely in that of cone $R$, hence no damping results from
particle-hole excitations of either cones here. 
The picture is dramatically different for cone $R$: its forbidden region overlaps with the damped region of 
cone $L$, i.e.,
the UHM of cone $R$ is strongly damped by particle-hole excitations in cone $L$.
For this reason, the UHM of cone $R$ disappears entirely.
In the opposite direction the roles of $L$ and $R$ are, of course, interchanged.
Therefore, the interaction of the two cones leads to a strong direction-dependent damping,
which results in the complete suppression of the UHM of
one cone where the other has particle-hole excitations of high spectral weight.
This phenomenon, which we refer to as \textit{intervalley damping}, is also visible in the 
imaginary part of the dielectric function (not shown).

The linear magnetoplasmons of each cone are situated in the particle-hole continuum of their own PHES respectively,
where they are already damped. 
They do not overlap with each other in the $(\omega,q)$ plane in the shown direction $\theta=\theta_\text{tilt}$.
Therefore, we expect the dominance of the one with higher spectral weight at a particular $(\omega,q)$.
Figure \ref{egydirection1}(c) testifies the dominance of the linear magnetoplasmon mode of cone $R$ in this
direction, in agreement with its higher spectral weight in the single cone model. 
This explains the reappearance of the linear magnetoplasmons at larger momenta,
and the disappearance of the linear magnetoplasmons of cone $L$ for intermediate momenta.

\begin{figure}[htbp]
\begin{center}
\includegraphics[width=\columnwidth, keepaspectratio]{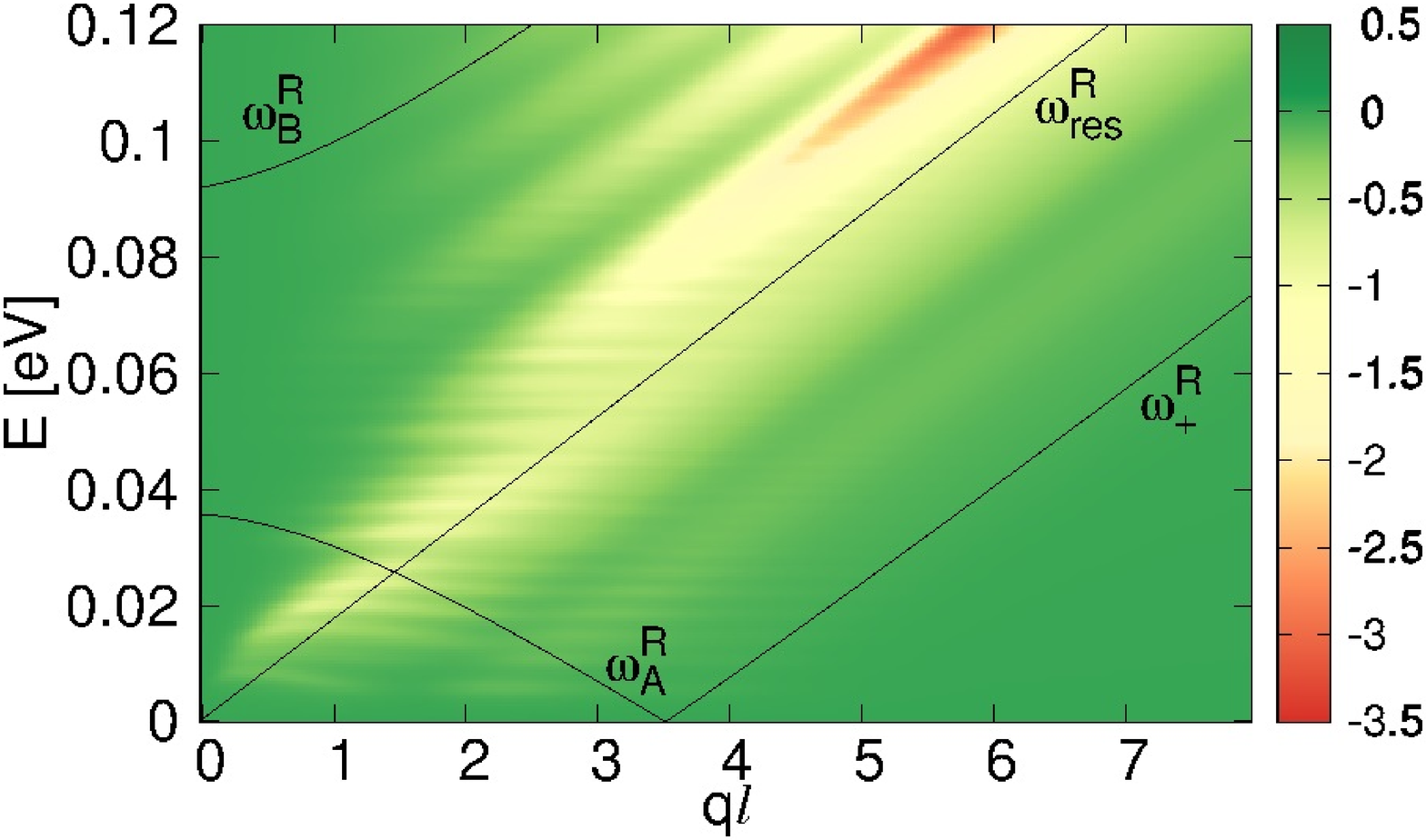}

\includegraphics[width=\columnwidth, keepaspectratio]{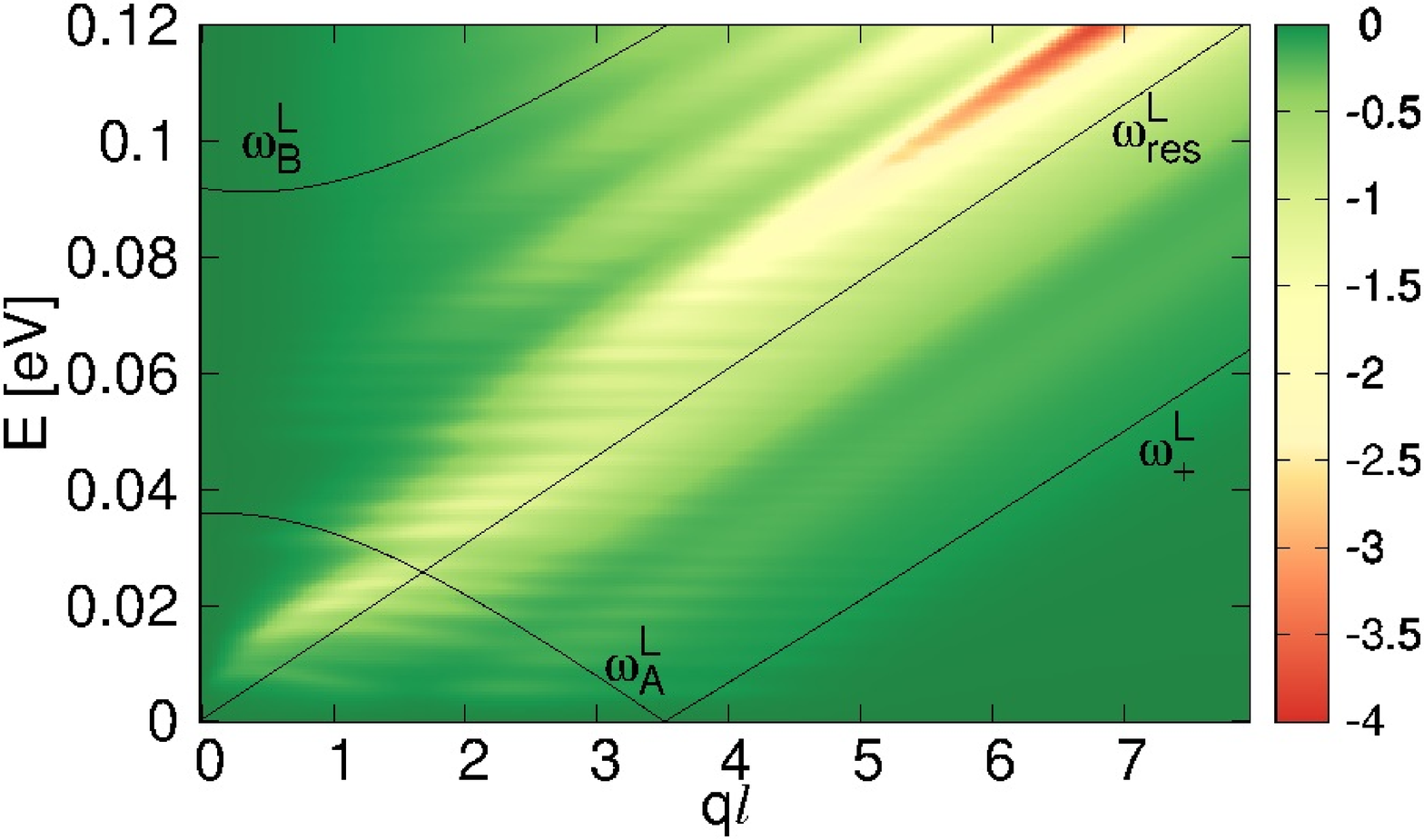}

\includegraphics[width=\columnwidth, keepaspectratio]{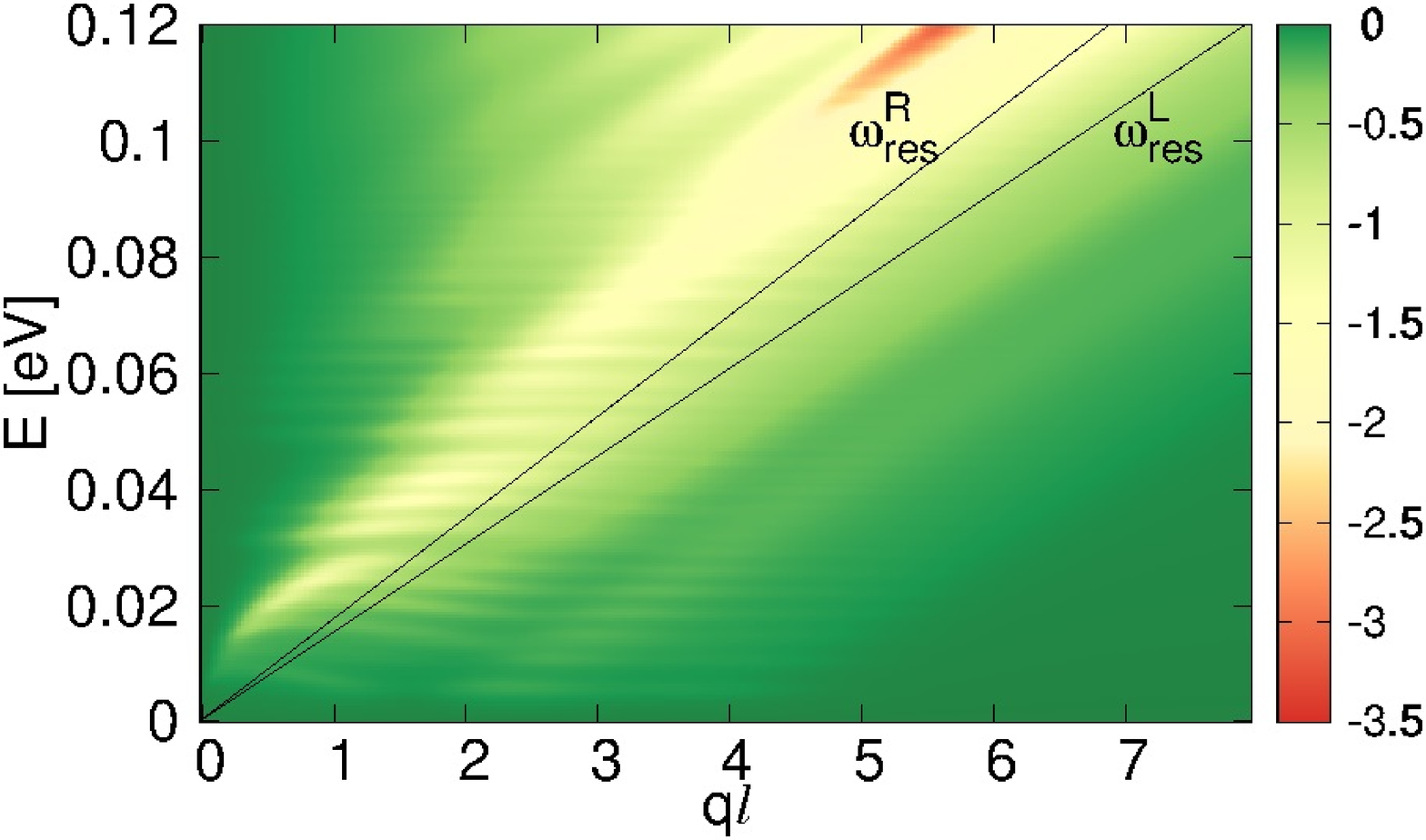}
\end{center}
\caption{
(Color online)
We show (a) $\chi^\text{RPA}_{R}$, (b) $\chi^\text{RPA}_{L}$, (c) $\chi^\text{RPA}_{L+R}$ in a fixed direction
$\theta=\theta_\text{tilt}+\pi/2$ on the momentum plane.
$B = 4$ T, $n_L^F = 2$, $\epsilon_r=10$. The lines denote the same boundaries as in Figs.~\ref{nishinetiltdirection} and \ref{egydirection1}.
}
\label{egydirection3}  
\end{figure}

Figure \ref{egydirection3} shows the density-density response in the direction $\theta=\theta_\text{tilt}+\pi/2$, i.e.,
perpendicular to the direction of the minimal steepness of cone $R$ (maximal steepness of cone $L$).
The response of the individual cones are almost identical [panels (a) and (b)], and their forbidden regions practically
coincide (perfect coincidence occurs at a nearby angle).
Intervalley damping is therefore absent in this direction.
The UHM and the linear magnetoplasmon mode of the two-valley system [panel (c)] are where they would be for
a single cone; albeit with an increased amplitude.

In order to illustrate the phenomenon of intervalley damping in a more precise and quantitative manner,
consider the bare polarizability (\ref{chi0teljes}) of a multivalley model, which is generically written as
\begin{equation}
\chi^{(0)}(\bq,\omega)=\sum_{V}\chi_V^{(0)}(\bq,\omega),
\end{equation} 
where the sum runs over all different valleys. One may thus rewrite the RPA dielectric function (21) as
\begin{equation}\label{eq:epsRPA}
\epsilon^{\rm RPA}(\bq,\omega)=\epsilon_{V_0}^{\rm RPA}(\bq,\omega) -v(\bq)\sum_{V\neq V_0}\chi_V^{(0)}(\bq,\omega),
\end{equation}
where $\epsilon_{V_0}^{\rm RPA}(\bq,\omega)$ is the RPA dielectric function within a single-valley model, where
only the valley $V_0$ is taken into account. The single-valley model therefore yields a good approximation 
for the collective modes [given by the zeros of $\epsilon^{\rm RPA}(\bq,\omega)$] if 
\begin{equation}
v(\bq)\sum_{V\neq V_0}\chi_V^{(0)}(\bq,\omega)\simeq 0
\end{equation}
in the region of interest, i.e. for $\omega=\omega_{pl}(\bq)$ obtained from the solution
$\epsilon_{V_0}^{\rm RPA}(\bq,\omega_{pl})=0$. This precisely means that the spectral weight for particle-hole
excitations in the other valleys $V\neq V_0$ vanishes in this region, 
or else that the collective modes of $V_0$ survive in
the forbidden regions of the other cones, as observed in our calculations. 

Notice that the phenomenon of intervalley damping is absent if all individual polarizabilities $\chi_V^{(0)}(\bq,\omega)$
are identical, e.g., in the absence of a tilt of the (albeit anisotropic) Dirac cones. In this case, the RPA
dielectric function (\ref{eq:epsRPA}) simply becomes 
\begin{equation}
\epsilon^{\rm RPA}(\bq,\omega)=1 - gv(\bq)\chi_V^{(0)}(\bq,\omega),
\end{equation}
for $g$ identical valleys.
Therefore, $\epsilon^{\rm RPA}(\bq,\omega)$ has only a single zero, at a slightly 
larger frequency as compared to the single-valley approximation because of the enhanced coupling 
$v(\bq)\rightarrow gv(\bq)$.
This situation is encountered, e.g., when one takes into account the spin degeneracy 
in conventional electron systems ($g=2$) or in graphene with non-tilted Dirac cones with a fourfold
spin-valley degeneracy ($g=4$).
The tilt of the Dirac cones, or more generally the broken $\bq\rightarrow -\bq$
symmetry in a single valley, is thus the basic ingredient for the mechanism of intervalley damping.

Finally, we note that Nishine \textit{et al.}\cite{Nishine} saw the effect of intervalley 
damping at zero magnetic field,
interpreted it as plasmon filtering,\cite{filtering} and offered a detailed analysis of the
angular dependence of strength of the lower-lying plasmon mode.
The two mechanisms are essentially the same, despite the slightly different formulations.

\subsection{Three-valley model}
\label{threevalleys}

Undoped samples with $B<B_{00}$ and hole-doped samples allow us to study the contribution of the quadratic valley.
In this subsection we investigate how this third valley damps the collective modes of the linear valleys,
and how the collective mode of the massive valley appears on alongside the excitations of the massless valleys.

The particle-hole continuum of the massive valley is bounded by\cite{2DEGchi}
\begin{equation}
\label{demarcate}
\omega^Q_\pm(\mathbf q)=\frac{\hbar^2q^2}{2m_Q}\pm \frac{\hbar^2qk^F_Q}{m_Q}.
\end{equation}
In the same manner as for the linear bands,
a 2D quadratic band has a plasmon mode that disperses $\propto\sqrt{q}$, and
in a magnetic field, this mode is becomes the gapped UHM.\cite{2DEGchi,chiuplasmon}

\begin{figure}[ht!]
\begin{center}
\includegraphics[width=\columnwidth,keepaspectratio]{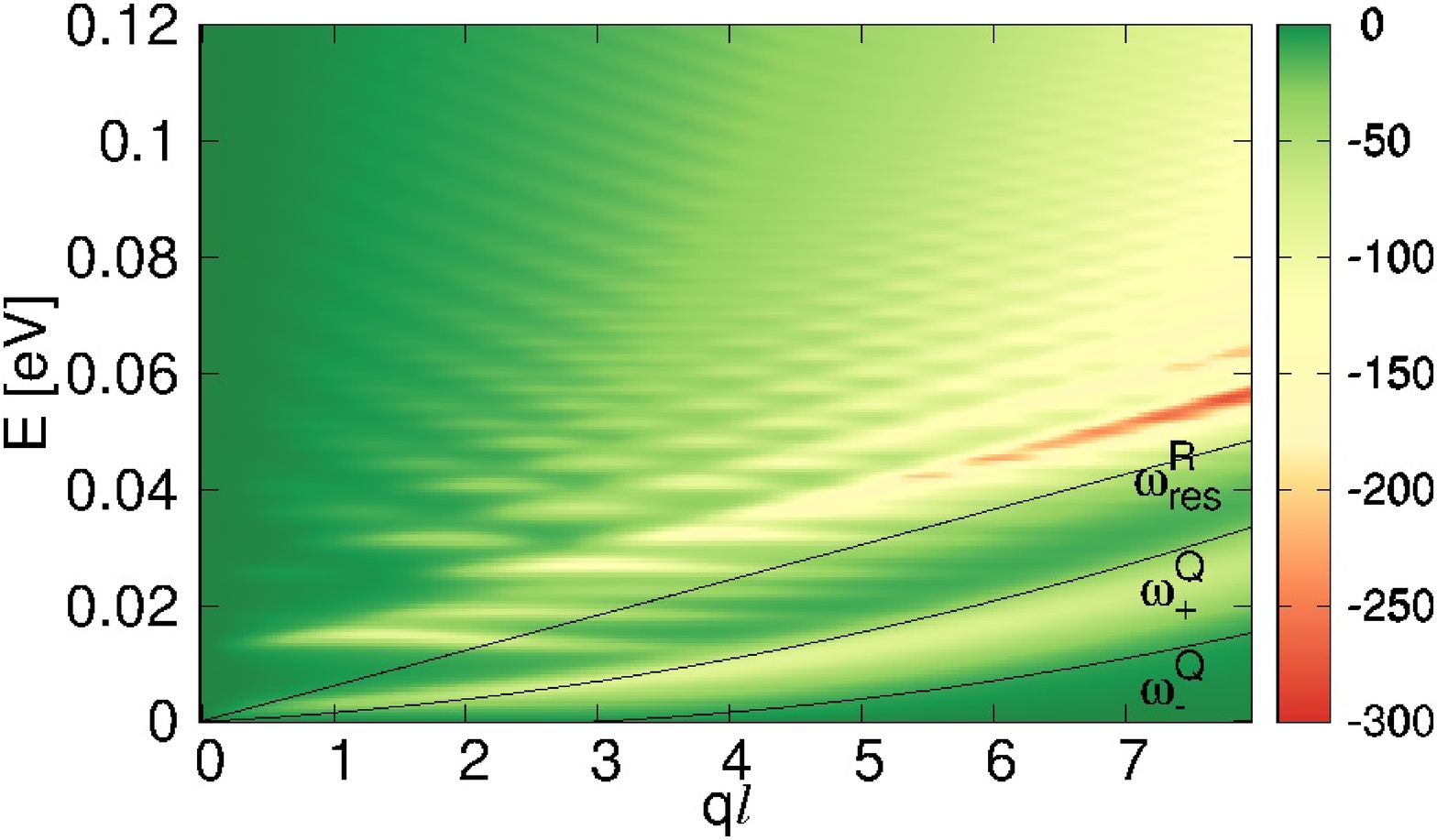}

\includegraphics[width=\columnwidth,keepaspectratio]{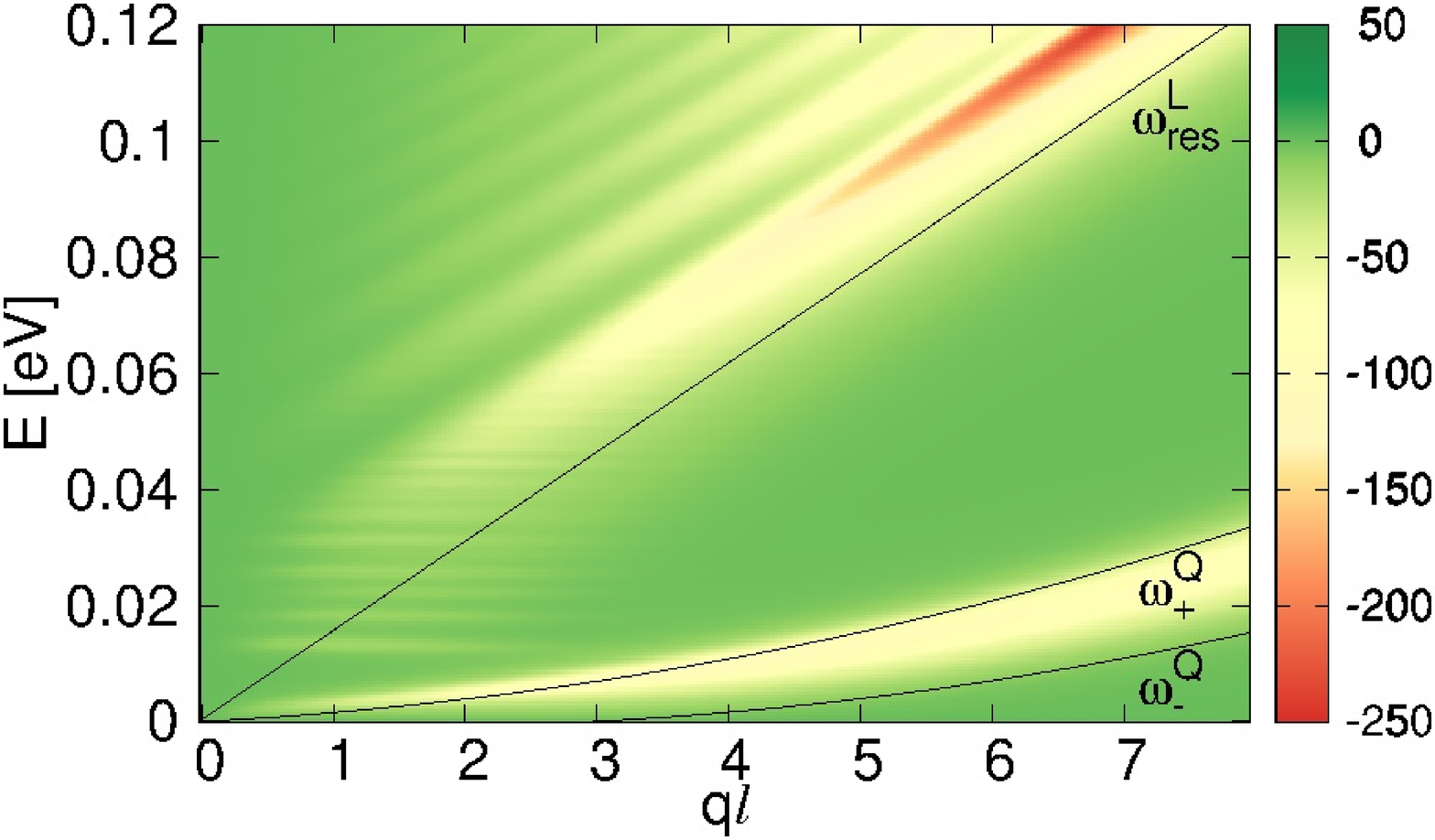}

\includegraphics[width=\columnwidth,keepaspectratio]{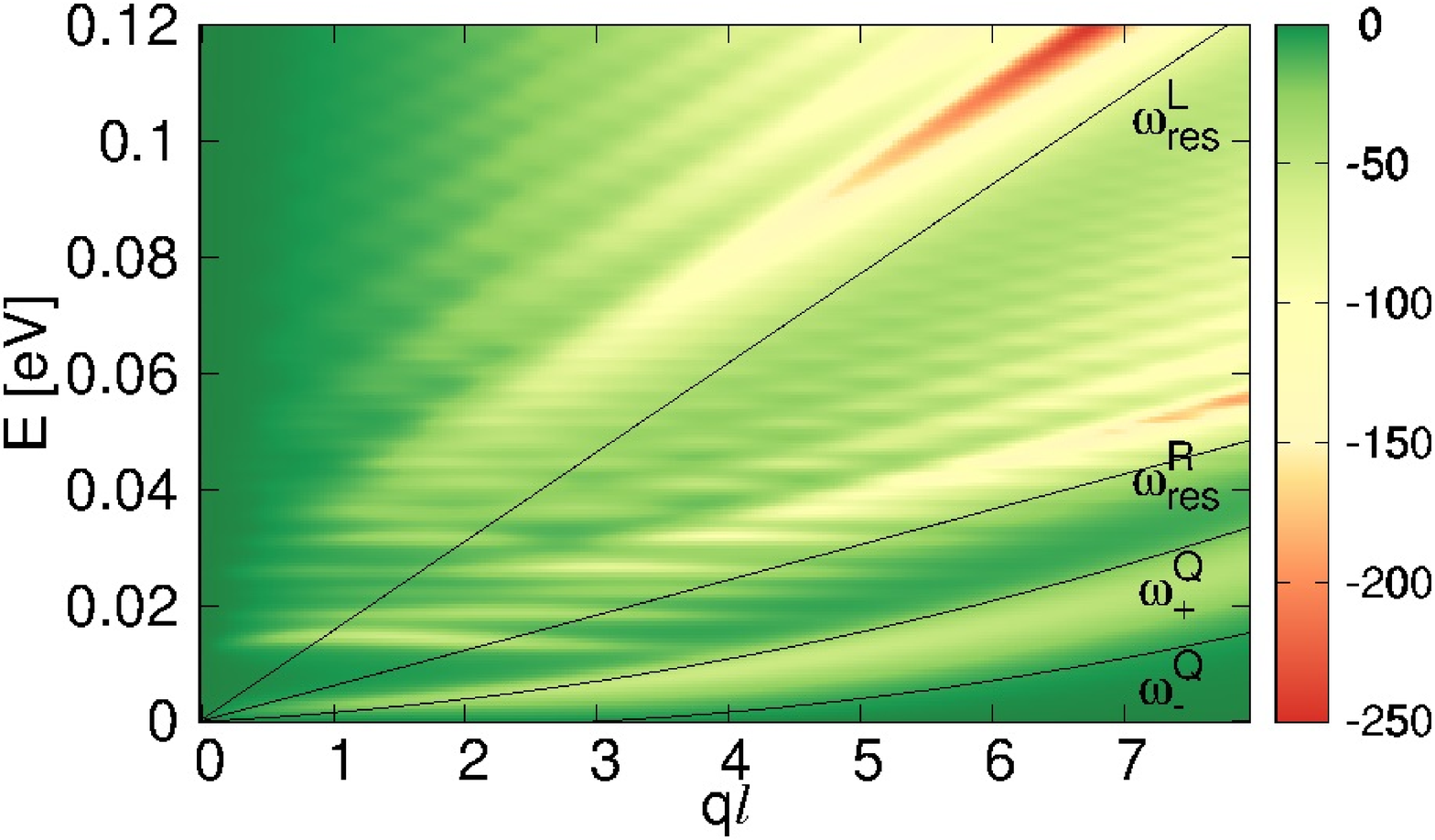}
\end{center}
\caption{
(Color online)
The density-density response in direction $\theta=\theta_\text{tilt}$ in momentum space.
The topmost filled band is $n_L^F = 0 $ in the massless valleys and $n_{Q}^F = 1$ in the massive valley.
Other parameters are $B = 2$ T and $\epsilon_r=10$.
Panel (a) shows the $\chi^\text{RPA}_{R+Q}$ of an imaginary system that omits cone $L$,
(b) shows $\chi^\text{RPA}_{L+Q}$ omits cone $R$, while (c) shows the complete $\chi^\text{RPA}$.
The units are s/m$^2$; normalization by the density of states is not applied.
The straight line $\omega_\text{res}^{R/L}$ is the boundary between the interband and intraband excitations of massless carriers,
and the curved ones [$\omega^Q_\pm$ in Eq.~(\ref{demarcate})] demarcate the particle-hole
continuum of massive carriers for $B=0$.
}
\label{massdirection1}  
\end{figure}

\begin{figure}[ht!]
\begin{center}
\includegraphics[width=\columnwidth,keepaspectratio]{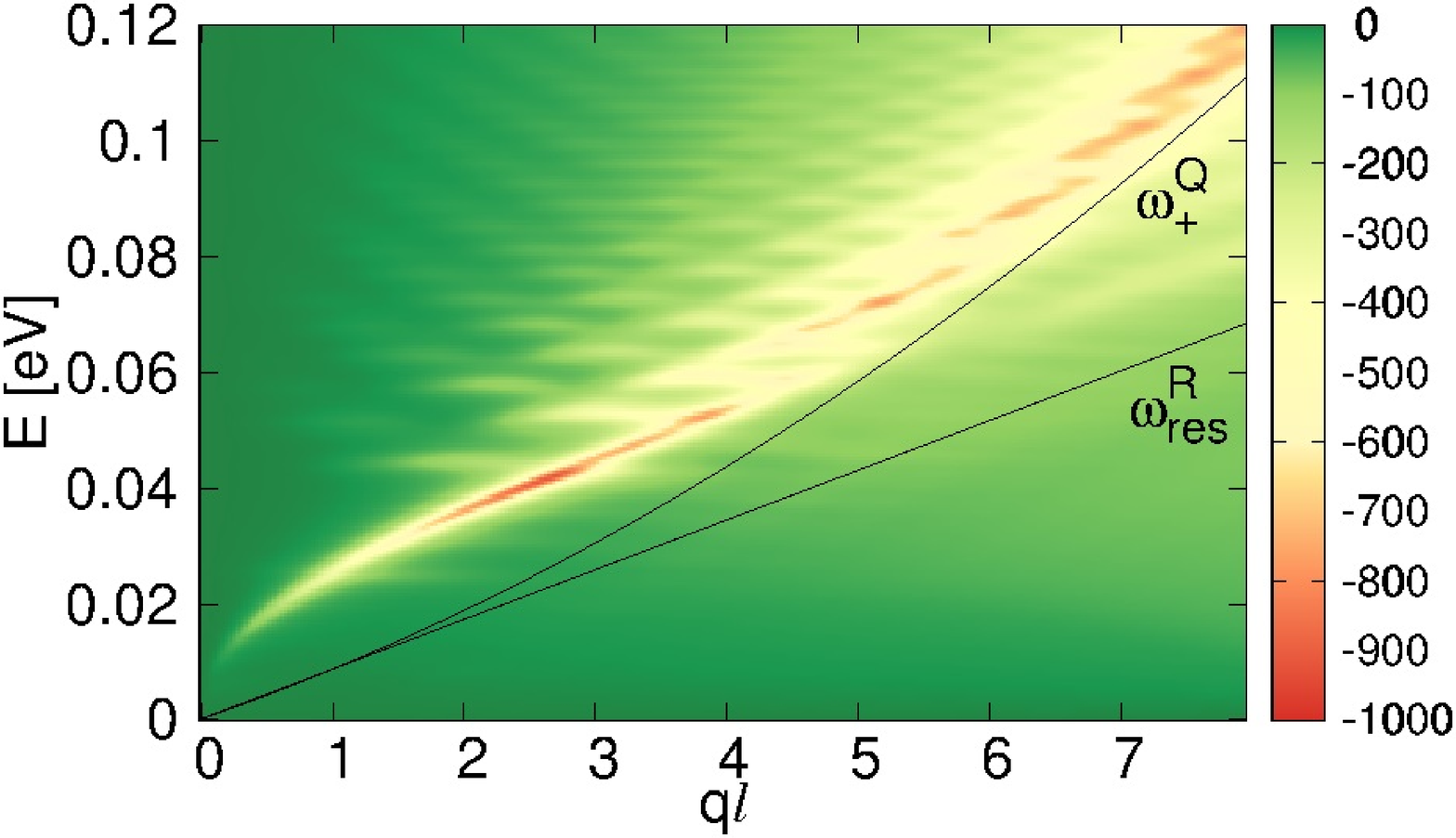}

\includegraphics[width=\columnwidth,keepaspectratio]{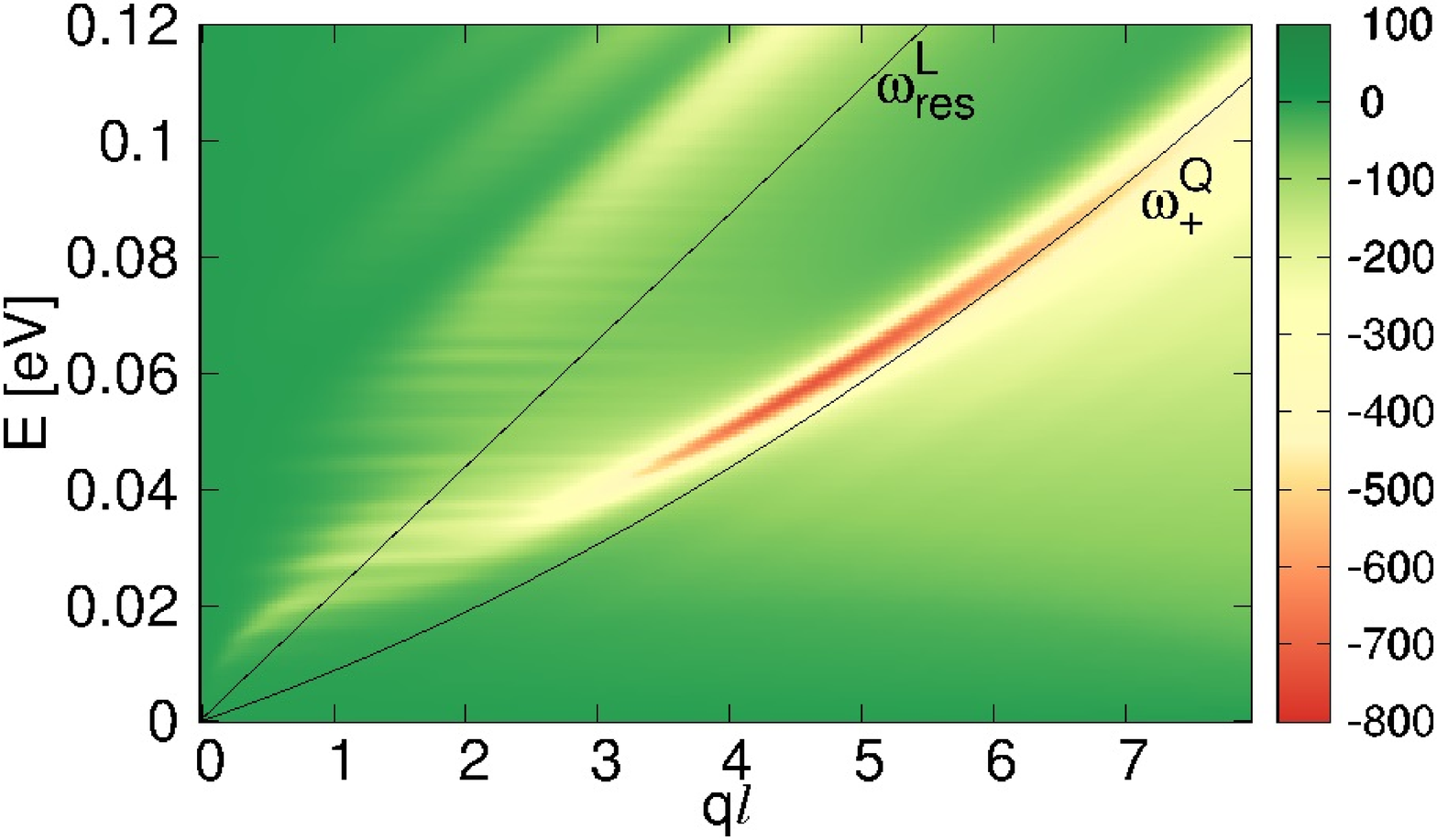}

\includegraphics[width=\columnwidth,keepaspectratio]{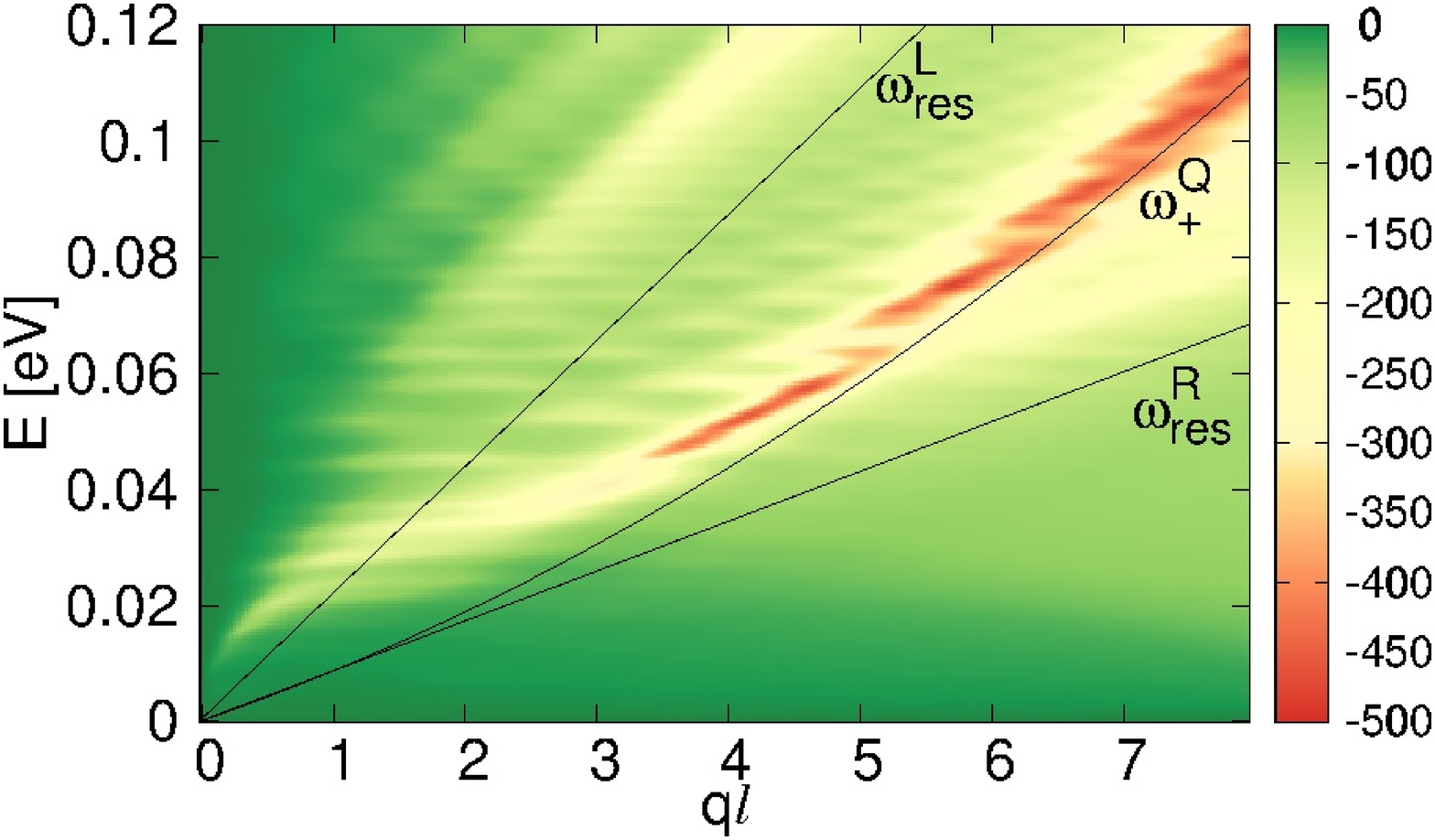}
\end{center}
\caption{
(Color online)
The same as Fig.~\ref{massdirection1}, but in a hole-doped sample.
The topmost filled massless Landau level is $n_L^F = -2$,
while among the massive LLs it is $n_{Q}^F = 13$.
Other parameters are $B = 4$ T and $\epsilon_r=10$.
For notations c.f.\ Fig.~\ref{massdirection1}.
Notice that one of the curves demarcating the particle-hole continuum of the massive valley for $B=0$,
$\omega^Q_-$ in Eq.~(\ref{demarcate}), is not visible because $2k^F_Q$ is outside of the presented momentum range.
}
\label{massdirection2}  
\end{figure}

Notice that the separation of the LLs scales as $\epsilon_{L,n}\propto\sqrt{B/n}$ and
$\hbar\omega_c\propto B$ for the massless and the massive carriers, respectively. One would therefore 
expect naively that the  LLs of the quadratic valley are generically less dense in energy than those of the Dirac cones.
However, due to the values for the Fermi velocities of the Dirac carriers and for the band mass of the massive
carriers, the LL separation of the massive carriers is much lower than that in the tilted Dirac cones in the magnetic-field
range discussed here. 
Furthermore, the relative position of the collective modes of the massive and the massless valleys and the
particle-hole continua of the three valleys are sensitive to both the magnetic field $B$ and the doping level.
Hence we discuss two representative cases, one at charge neutrality and relatively small magnetic fields $B<B_{00}$
(Fig.~\ref{massdirection1}), and one at heavy hole-doping (Fig.~\ref{massdirection2}). The two situations correspond
to the sketches in Figs. \ref{landaulevels}(b) and (c), respectively.

The first case is studied in Fig.~\ref{massdirection1}, which depicts the density-density response
at $n_L^F=0$ and $n_Q^F=1$, which is adequate at charge neutrality when $B_{11}<B<B_{00}$; c.f.\ Fig.~\ref{landaulevels}(b).
Now the particle-hole continuum of the quadratic valley lies
well below the linear magnetoplasmon mode of
the massless Dirac carriers in both cones in the direction $\theta=\theta_\text{tilt}$,
for which the difference between the velocities in the two cones is the most pronounced.
This is demonstrated in Fig.~\ref{massdirection1}(a,b), where one of the cones
are disregarded for visibility reasons. In addition to the characteristic frequency $\omega_{res}^{R/L}$, we have
also sketched the lines $\omega_{+}^Q(\mathbf q)$ and $\omega_{-}^Q(\mathbf q)$, which delimit the particle-hole
continuum of the massive carriers. 
Thus the particle-hole excitations of the massive carriers do not overlap with those of the Dirac valleys,
and its spectral weight is comparable to the dominant linear magnetoplasmon modes of the latter.
The UHM of the massive holes is not damped either, as it lies at very small energies,
well below the cyclotron frequency of the massless carriers.
(The former is characterized by the energy scale $\hbar\omega_c\approx0.00075$ eV,
while the latter is given by $\epsilon_{L,1}\approx0.015$ eV at $B=2$ T.)
The total, physical density-density response in Fig.~\ref{massdirection1}(c) indicates that the response of
$\alpha-$(BEDT-TTF)$_2$I$_3$ at low frequencies is determined by the massive carriers. In the higher frequency range
the Dirac carriers dominate, with intervalley-damped collective modes, as discussed in \S\ref{bothcones} above,
but no damping due to the massive carriers.
The reason is the separation of energy scales at low momenta, and not any difference in the density of states.

Figure \ref{massdirection2} shows the second case, a strongly hole-doped situation with $n_L^F=-2$ and  $n_Q^F=13$.
Now the massive carriers are definitely involved in the damping.
In direction $\theta=\theta_\text{tilt}$,
the forbidded region 1B of cone $R$ overlaps with
the particle-hole continuum of the massive valley [Fig.~\ref{massdirection2}(a)].
The UHM, which stems from the Dirac carriers, is still obtained at larger energies than 
that associated with the massive carriers. 
(At $B=4$ T the cyclotron frequency is characterised by the energy scale $\hbar\omega_c\approx0.0015$ eV for the massive carriers,
while by $\epsilon_{L,1}\approx0.02$ eV for the massless ones.)   
The latter is clearly discernible in Fig.~\ref{massdirection2}(a),
and it is damped once it enters the overlap region between the two lines 
$\omega_+^Q$ and $\omega_{res}^R$, which is not a forbidden region for particle-hole excitations of the massive carriers. Furthermore, it
is clear from Fig.~\ref{massdirection2}(a) that the UHM avoids entering this region and rather approaches asymptotically the border 
$\omega_+^Q$, as one expects for the UHM in a single-band model of massive carriers.\cite{2DEGchi,chiuplasmon} However, the situation
is slightly more involved in the present case, where one observes a coupling between the UHM with the linear magnetoplasmons of the massless
Dirac carriers, which gives rise to a modulation of the spectral weight along the UHM. This coupling is reminiscent to the so-called Bernstein modes
discussed in the framework of graphene in a strong magnetic field.\cite{BernsteinG}
For cone $L$ (or cone $R$ in the $\theta=\theta_\text{tilt}+\pi$ direction), on the other hand,
the forbidden region of the massive valley overlaps with the 1A intraband region of the massless valley.
The UHM is therefore strongly damped and almost invisible at small momenta but
reappears at $q\ell\gtrsim 3$, c.f. Fig.~\ref{massdirection2}(b).
This reappearance of the UHM, even if it occurs in a region of possible intraband particle-hole excitations of the massless Dirac carriers, 
may be understood from the particular chiral properties of the latter. Indeed, the spectral weight of particle-hole excitations, in the case of 
Dirac fermions, is concentrated around the lines $\omega_{res}$ due to the suppression of 
backscattering,\cite{grafenBnull,grafenB,goerbigsummary,KotovDrut} such that the UHM is barely damped once it is further away
from $\omega_{res}$.

In the regions where the allowed regions (particle-hole continua) overlap, the one with
the greater spectral weight dominates.
This is visible in Fig.~\ref{massdirection2}(a), where the particle-hole continuum of the massive
valley is almost as bright as the more concentrated linear magnetoplasmon mode of the Dirac fermions.
This also applies to the total, physical density-density response, Fig.~\ref{massdirection2}(c),
where the spectral weight is mostly concentrated on the excitations of the massive valley,
although the linear magnetoplasmon of cone $R$ become visible at large momenta.

\subsection{Static screening}

\begin{figure}[htbp]
\begin{center}
\includegraphics[height=\columnwidth,angle=270,keepaspectratio]{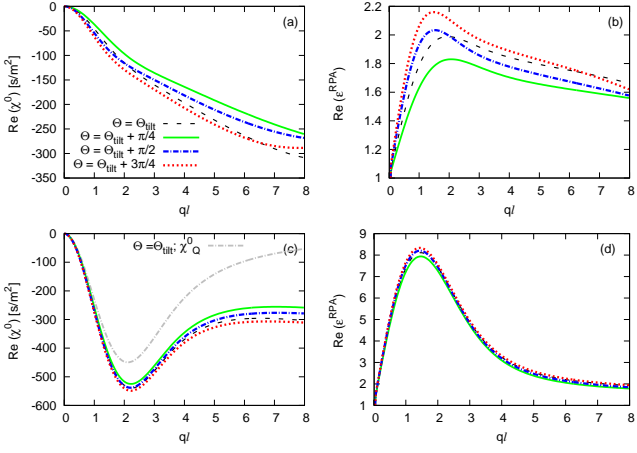}
\end{center}
\caption{\label{static}
(Color online)
The static screening properties of $\alpha-$(BEDT-TTF)$_2$I$_3$ at charge neutrality.
Upper row: $B=4$ T, the massive valley is completely filled and inert.
Lower row: $B=2$ T, the topmost Landau level of the massive valley is empty.
Left panels: $\text{Re}\chi^{(0)}(\mathbf q,0)$ in s/m$^2$, right panels: $\text{Re}\epsilon^\text{RPA}(\mathbf q,0)$.
We show several directions in the momentum plane, specified by the angle $\theta$ relative to $\theta_\text{tilt}$,
the tilting direction of cone $R$.
The grey line in panel (c) shows the polarizability of the massive valley, which is dominant and
suppresses any anisotropy due to the tilted Dirac cones if $B<B_{00}$.
}
\end{figure}

The interplay between massless carriers in tilted anisotropic Dirac cones and massive holes in a roughly isotropic
pocket gives rise to a remarkable doping-dependent angular dependence of the screening properties.
Fig.~\ref{static} shows the real part of the static bare polarizability
$\text{Re}\chi^{(0)}(\mathbf q,0)$, and the dielectric function
$\text{Re}\epsilon^\text{RPA}(\mathbf q,0)$ in the RPA, respectively.
For electron-doping or charge neutrality at $B>B_{00}\approx2.5$ T,
the bare polarizability is entirely due to the linear bands [panel (a)].
This anisotropy is naturally inherited by the RPA dielectric function [panel (b)],
which means that screening due to the discussed bands is anisotropic.
If the magnetic field is reduced, $B<B_{00}$, however, massive carriers can also be polarized at zero temperature.
Because of their higher density of states at small energies, their contribution is dominant in the total
polarizability, which shows a very weak angular dependence [panel(c)].
As a result, screening is almost isotropic; c.f.\ Fig.~\ref{static}(d).
The suppression of the screening anisotropy of the tilted cones by the massive valley naturally
becomes stronger in hole-doped samples.

\section{Conclusion}
\label{conclu}

In this work we have studied the low-energy magnetic excitations of the quasi-2D electron gas
in a three-valley system of two tilted and anisotropic massless Dirac cones and a massive hole pocket,
with an emphasis on direction-dependent effects and the properties for which massive carriers are relevant.
This model is a realistic representation of a layer of the organic conductor $\alpha-$(BEDT-TTF)$_2$I$_3$ at high pressure
or uniaxial strain, but there is some ambiguity regarding the band parameters and even the presence of massive carriers.
We have found that the tilt of the cones causes a direction-dependent intervalley damping of the
upper hybrid modes of the Dirac valleys, while the linear magnetoplasmons are less affected.
The magnetoplasmons of the massive band may coexist with those of the massless ones,
depending on doping and the strength of the magnetic field.
The latter also tunes the system between isotropic and anisotropic screening regimes.

The experimental study of the collective modes in ribbon samples\cite{Poumirol} or by local probes,\cite{Fei}
successfully applied for graphene, may be challenging under high pressure, but grating couplers might be usable.
On the other hand, the layered structure of these organic conductors recommends itself to inelastic light
scattering experiments,\cite{layeredraman} which have successfully clarified the intrasubband plasmon modes
of multilayer GaAs-(AlGa)As heterostructures.
The elaboration of collective modes in the presence of interlayer coupling is delegated to future work.

\section*{Acknowledgements}
This research was funded by the Hungarian Scientific Research Funds No.\ K105149.
C.\ T.\ was supported by the Hungarian Academy of Sciences.
Supercomputer facilities were provided by National Information Infrastructure Development Institute, Hungary.
J.\ S.\ was supported by Campus France and acknowledges hospitality from Laboratoire de Physique des Solides.

\appendix

\section{Linear valleys in a magnetic field}
\label{linorbit}

Here follow the Morinari \textit{et al.}'s Ref.~\onlinecite{Morinari2009}, merely fixing some apparent bugs in the formulas.
We use the Landau gauge $\mathbf{A} = (-yB,0,0)$, and work in the coordinate system introduced in Eq.~(\ref{newcoord}).
The transformed real-space coordinates $\mathbf{\tilde r}=(\tilde x,\tilde y)$ are
\begin{equation}
\label{newcoord2}
\left.\begin{array}{l}
\tilde{x} = x \cos\theta +  \alpha^2  y\sin\theta\\
\tilde{y} = -x\sin\theta + \alpha^2 y\cos\theta
\end{array}\right\}.
\end{equation}

For $n \neq 0$ in cone $R$ the Landau orbitals are
\begin{multline}
\Phi_{R,n,k}(\mathbf{\tilde r}) =
\frac{1}{\sqrt{4 (1 + \lambda)}}
\left[  \begin{pmatrix}  -\eta  \\ 1+ \lambda  \end{pmatrix} \phi_{R,n,k}(\mathbf{\tilde r})+ \label{nagypszi}\right.
\\
+\left.\begin{pmatrix}-1-\lambda \\ \eta\end{pmatrix}  \text{sgn}(n)\phi_{R,n-1, k} (\mathbf{\tilde r}) \right],
\end{multline}
while for the $n= 0$ Landau level,
\begin{equation}
\Phi_{R,0,k}(\mathbf{\tilde r})=
\frac{1}{\sqrt{2 (1 + \lambda)}}\ \begin{pmatrix} - \eta\\ 1+ \lambda  \end{pmatrix} \phi_{R,0,k}(\mathbf{\tilde r}) .
\end{equation}
For $n \neq 0$ in cone $L$, the orbitals are
\begin{multline}
\label{nagypszil}
\Phi_{L,n,k}(\mathbf{\tilde r})=
\frac{1}{\sqrt{4 (1 + \lambda)}}\left[\begin{pmatrix}1+ \lambda \\ -\eta\end{pmatrix}\phi_{L,n,k}(\mathbf{\tilde r})\right.+
\\
+\left.\begin{pmatrix}-\eta \\1+\lambda \end{pmatrix}  \text{sgn}(n)\phi_{L,n-1, k} (\mathbf{\tilde r}) \right],
\end{multline}
while for the $n=0$,
\begin{equation}
\Phi_{L,0,k}(\mathbf{\tilde r})  =
\frac{1}{\sqrt{2 (1 + \lambda)}}\ \begin{pmatrix} -1-\lambda \\  \eta\end{pmatrix} \phi_{L,0,k}(\mathbf{\tilde r}) .
\label{nagypszivege}
\end{equation}

We have used the subformulas ($\xi\in\{R,L\}$)
\begin{equation*}
\phi_{\xi,n, k} (\mathbf{\tilde r})  =  \frac{\lambda^{1/4}}{\sqrt{2\pi}} \frac{e^{i k\tilde{ x}}}{(2^{|n|} |n|! \sqrt{\pi} \alpha \ell)^{1/2}}  e^{-\textsc{y}^2_{n, \xi} /2} H_{|n|} (\textsc{y}_{n, \xi} ) 
\end{equation*}
and
\begin{multline*}
\phi_{\xi,n-1, k} (\mathbf{\tilde r})=\\
\frac{\lambda^{1/4}}{\sqrt{2\pi}} \frac{e^{i k \tilde{x}}}{(2^{|n-1|} |n-1|! \sqrt{\pi}\alpha \ell)^{1/2}}  e^{-\textsc{y}^2_{n, \xi} /2 } H_{|n-1|} (\textsc{y}_{n, \xi}),
\end{multline*}
with the argument 
\begin{equation*}
\textsc{y}_{n,R/L}=
\sqrt{\lambda}\tilde{y}/ \alpha \ell  -\sqrt{\lambda} \alpha \ell k\mp\eta \sqrt{2 |n|} \text{sgn}(n).
\end{equation*}
Notice that $\textsc{y}_{n,\xi}$ depends on both the LL index (including its sign) and the cone $\xi$.
Also, it is identical for the $\phi_{\xi,n, k}$ and $\phi_{\xi,n-1, k}$ parts of $\Phi_{\xi,n, k}$ in Eqs.~(\ref{nagypszi})
and (\ref{nagypszil}).

\section{Bare polarizability of the linear valleys}
\label{applinpol}

Here we calculate the bare polarizability of the linear valleys by standard methods, using the orbitals that
Eqs.~(\ref{nagypszi}) to (\ref{nagypszivege}) specify in the rotated, rescaled coordinate system,
c.f.\ Eqs.~(\ref{newcoord}) and (\ref{newcoord2}).

The field operators are ($\xi\in\{R,L\}$):
\begin{equation}
\Psi_\xi (\mathbf{r}, t ) = \sum_n \int \mathrm{d}q  \Phi_{\xi,n,k}(\mathbf{r}) e^{-i t\epsilon_{L,n}} c _{\xi,n,q}.
\end{equation}
Suppressing spin to avoid clutter, the (gauge-dependent) bare Green's function has a matrix structure,
\begin{multline}
\mathcal{G}^{(0)}_\xi(\mathbf{R}, \Delta\mathbf{r}, t) = \\
= -i\left\langle \mathcal T \Psi_\xi\left(\mathbf{R}+\frac{\Delta \mathbf{r}}{2}, t\right)\otimes
\Psi^\dag_\xi \left(\mathbf{R}-\frac{\Delta \mathbf{r}}{2}, 0\right)\right\rangle,
\end{multline}
where $\mathbf R=(\mathbf r+\mathbf r')/2$ and $\Delta\mathbf r=\mathbf r-\mathbf r'$.
By Fourier-transformation,
\begin{multline}
\mathcal{G}^{(0)}_\xi(\mathbf{R}, \mathbf{p}, E)
=\sum_{n}\int \mathrm{d}q\int\mathrm{d}^2\Delta\mathbf{r}e^{i\mathbf{p}  \Delta \mathbf{r}}\times\\
\times\frac{ \Phi_{\xi,n,q}(\mathbf{R}+\frac{\Delta \mathbf{r}}{2})
\otimes\Phi^\dag_{\xi,n,q}(\mathbf{R}-\frac{\Delta \mathbf{r}}{2})}
{E-\epsilon_{L,n} +i \eta\text{sgn}(\epsilon_{L,n} -\epsilon^F_L)}.
\end{multline}

Now we can evaluate the bare polarizability in Eq.~(\ref{chidef}),
involving the matrix structure of the Green's functions in the trace.
We obtain Eq.~(\ref{chi0lin}), with the form factors $\mathcal{F}^{\xi}_{n,n'}$ defined as
\begin{multline}
\label{form1}
\mathcal{F}^{\xi}_{n', n}(\mathbf{\tilde q})  =  \frac{1}{2} F_{|n'|,|n|} ^{n', n, \xi } (\mathbf{\tilde q})
+\frac{1}{2}\text{sgn}(n) \text{sgn}(n') F_{|n'|-1,|n|-1}^{n', n, \xi} (\mathbf{\tilde q})\\
+ \frac{ \xi \eta }{2} \text{sgn}(n) F_{|n'|,|n|-1} ^{n', n, \xi} (\mathbf{\tilde q})
+ \frac{\xi \eta}{2} \text{sgn}(n') F_{|n'|-1,|n|} ^{n', n, \xi } (\mathbf{\tilde q})
\end {multline}
for $n, n' \neq 0$.
Similarly for $n\neq 0=n'$,
\begin{multline}
\mathcal{F}^{\xi}_{0, n}(\mathbf{\tilde q})=
\frac{1}{\sqrt{2}} F_{0, |n|}^{0, n, \xi} (\mathbf{\tilde q})+
\frac{\xi  \eta }{\sqrt{2}} \text{sgn}(n) F_{0,|n|-1} ^{0, n, \xi}(\mathbf{\tilde q}) ; 
\end{multline}
and finally, for $n=n'=0$,
\begin{eqnarray}
\mathcal{F}^{\xi}_{0, 0}(\mathbf{\tilde q}) & = &  F_{0,0}^{0, 0, \xi} (\mathbf{\tilde q}).
\end {eqnarray}

Here we have introduced the functions $F_{|n'|,|n|}^{n', n, \xi} (\mathbf{\tilde q})$, of
$\mathbf{\tilde q}=(\tilde q_x,\tilde q_y)$.
For $|n'| \geq |n|$ they are defined as  
\begin{multline*}
F_{|n'|,|n|}^{n', n, \xi}(\mathbf{\tilde q}) =
\sqrt{\frac{|n|!}{|n'|!}}\sqrt{2}^{|n'|-|n|} \ (- i P +Q_{ n',n, \xi})^{|n'|-|n|}\times\\ 
\times L_{|n|}^{|n'|-|n|} (2 (  Q_{ n',n, \xi}^2 + P^2 )) e^{- (Q_{n',n, \xi}^2 + P^2)}.
\end{multline*}
Similarly, for $|n| > |n'|$ the definition is
\begin{multline}
F_{|n'|,|n|}^{n', n, \xi} (\mathbf{\tilde q})  =
\sqrt{\frac{|n'|!}{|n|!}}\sqrt{2}^{|n|-|n'|} \left(- i P - Q_{ n', n, \xi}\right)^{|n|-|n'|} \times\\ 
\times L_{|n'|}^{|n|-|n'|}\left(2(Q_{ n', n, \xi}^2 + P^2)\right) e^{- (Q_{ n', n, \xi}^2 + P^2)}.
\label{form2}
\end{multline}
In the above definitions
\begin{gather*}
Q_{ n',n, \xi}= \frac{\tilde{q}_x \sqrt{ \lambda} \alpha \ell - \xi \eta \sqrt{2 |n'|} \text{sgn}(n')+
\xi \eta \sqrt{2|n|}\text{sgn}(n)}{2},\\
P = \frac{\alpha \ell}{2\sqrt{\lambda}} \tilde{q}_y
\end{gather*}
It is easy to check that
$F_{|n'|,|n|} ^{n', n, \xi }(\mathbf{\tilde q})=\left[F_{|n|, |n'|} ^{n', n , \xi }(-\mathbf{\tilde q})\right]^\ast$.


\begin{thebibliography}{200}
\bibitem{footfilling} Notice that there are four BEDT-TTF molecules and two I$_3^-$ ions in a unit cell.
The latter capture two out of the eight electrons in the highest occupied molecular orbitals of the BEDT-TTF's,
which comprise the conducting layer.
\bibitem{holedoping}
Notice the tension arises because the quadratic valley has a maximum near the Fermi energy,
thus it has little effect if the sample is electron-doped, but as its set of Landau levels is denser
than that of the linear valleys, several of its LLs are emptied for hole-doping before the
first LLs of the linear valleys are affected.
\end{thebibliography}
\end{document}